\documentclass{llncs}
\pdfoutput=1
\usepackage{amssymb}
\usepackage{bm}
\usepackage{inputenc}
\usepackage{graphicx}
\usepackage{amsmath}
\begin{document}
\title{Semi-online Scheduling: A Survey}
\author{Debasis Dwibedy and Rakesh Mohanty}
\institute{Department of Computer Science and Engineering \\ Veer Surendra Sai University of Technology, Burla\\ Sambalpur, Odisha - 768018, India \\ debasis.dwibedy@gmail.com, rakesh.iitmphd@gmail.com}
\maketitle
%
%
\begin{abstract}
Scheduling of jobs on multiprocessing systems has been studied extensively since last five decades in two well defined algorithmic frameworks such as offline and online. In offline setting, all the information on the input jobs are known at the outset. Whereas in online setting, jobs are available one by one and each job must be scheduled irrevocably before the availability of the next job. Semi-online is an intermediate framework to address the practicability of online and offline frameworks. Semi-online scheduling is a relaxed variant of online scheduling, where an additional memory in terms of buffer or an \textit{Extra Piece of Information(EPI)} is provided along with input data. The \textit{EPI} may include one or more of the parameter(s) such as  size of the largest job, total size of all jobs, arrival sequence of the jobs, optimum makespan value or range of job's processing time. A semi-online scheduling algorithm was first introduced in 1997 by Kellerer et al. They envisioned semi-online scheduling as a practically significant model and obtained improved results for $2$-identical machine setting. 
This paper surveys scholarly contributions in the design of semi-online scheduling algorithms in various parallel machine models such as identical and uniformly related by considering job's processing formats such as preemptive and non-preemptive with the optimality criteria such as \textit{Min-Max} and \textit{Max-Min}. The main focus is to present state of the art competitive analysis results of well-known semi-online scheduling algorithms in a chronological overview. The survey first introduces the online and semi-online algorithmic frameworks for the multi-processor scheduling problem with important applications and research motivation, outlines a general taxonomy for semi-online scheduling. Fifteen well-known semi-online scheduling algorithms are stated. Important competitive analysis results are presented in a chronological way by highlighting the critical ideas and intuition behind the results. An evolution time-line of semi-online scheduling setups and a classification of the references based on \textit{EPI} are outlined. Finally, the survey concludes with the exploration of some of the interesting research challenges and open problems.  

\end{abstract}
\section{Introduction} 
\label{sec:Introduction}
Scheduling deals with allocation of resources to jobs in some order with application specific objectives and constraints. The concept of scheduling was introduced to address the following research question [1]: 
\textit{Given a list of $n$ jobs and $m$($\geq 2$)machines, what can be a sequence of executing the jobs on the machines such that all jobs are finished by latest time possible?}
Scheduling has now become ubiquitous in the sense that it inherently appears in all facets of daily life. Everyday, we involve ourselves in essential activities such as scheduling of meetings, setting of deadlines for projects, scheduling the maintenance periods of various tools, planning and management of events, allocating lecture halls to various courses, organizing vacations, work periods and academic curriculum etc. Scheduling finds practical applications in broad domains of computers, operations research, production, manufacturing, medical, transport and industries [17]. Widespread applicability has made scheduling an exciting area of investigation across all domains. \\
Scheduling of jobs on multiprocessing systems has been studied extensively over the years in well defined algorithmic frameworks of offline and online scheduling [6, 16, 17, 41, 49]. A common consideration in \textit{offline scheduling} is that all information about the input jobs are known at the outset. However, in most of the current practical applications, jobs are given incrementally one by one. An irrevocable scheduling decision must be made upon receiving a job with no prior information on successive jobs [2, 12]. Scheduling in such applications is known as \textit{online scheduling}. In this survey, we study a relaxed variant of online scheduling, known as \textit{semi-online scheduling},  where some \textit{extra piece of information} about the future jobs are known at the outset. We present the structure and organization of our survey in Figure 1.
\begin{figure}[h]
\centering
\includegraphics[scale=0.80]{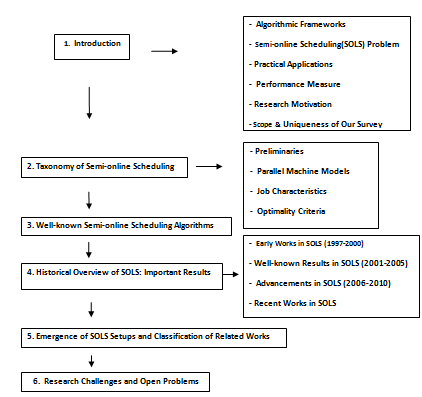}
\caption{Organization of our Survey}
\label{fig:structureofoursurvey.png}  
\end{figure}
\subsection{Algorithmic Frameworks}
\label{subsec:Algorithmic Frameworks}
\begin{figure}[h]
\centering
\includegraphics[scale=0.55]{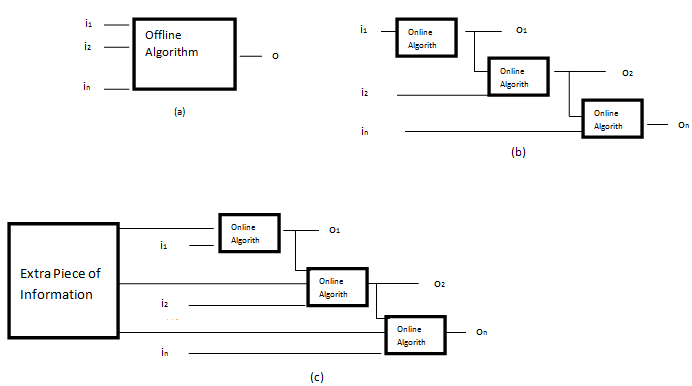}
\caption{Algorithmic Frameworks (a) Offline (b) Online  (c) Semi-online}
\label{fig:processingframeworks.png}  
\end{figure}

We present three algorithmic frameworks such as offline, online and semi-online based on availability of input information in processing of a computational problem as shown in Figure \ref{fig:processingframeworks.png}.
\begin{itemize}
\item
In \textbf{Offline framework}, complete input information is known at the outset. Let us consider a set $I$=$\{i_1,i_2,...,i_n\}$ representing all inputs of a computational problem $X$. In offline framework, $I$ is known prior to construct a solution for $X$. The algorithm designed for computation of $I$ in the offline framework is known as \textit{offline algorithm}. An offline algorithm processes all inputs $I$ simultaneously to produce the final output $o$. 
\item In \textbf{Online framework}, the inputs are given one by one in order. Each available input must be processed immediately with no information on the successive inputs. In online framework, at the time step $t$, the input sequence $I_t$:$<i_1,i_2,...,i_{t-1},i_t>$ is known and must be processed irrevocably with no information on future input sequence $<i_{t+1},...,i_{n-1},i_n>$, where $t\geq 1$. The algorithm designed for computation of $I$ in the online framework is known as \textit{online algorithm}. An online algorithm produces a partial output $o_t$ for each input $I_t$ on the fly, where, $1\leq t\leq n-1$ before producing the final output $o_n$. 
\item In \textbf{Semi-Online framework} inputs are given one by one like online framework along with some \textit{Extra Piece of Information(EPI)} on future inputs. At any time step $t$, a semi-online algorithm receives input sequence $I_t$ with an \textit{EPI} and processes them irrevocably to obtain a partial output $o_t$ on the fly, where $1\leq t\leq n-1$ before producing the final output $0_n$. 
\end{itemize} 
Semi-online is an intermediate framework to address the practicability and limitations of online and offline frameworks. In most of the current practical scenarios, neither all the inputs are available at the beginning nor the inputs occur exclusively in online fashion, but may occur one by one with additional information on the successive inputs. For example, an online \textit{video on demand} application receives requests for downloading video files on the fly, however, it knows the highly requested video file and the largest video file among all video files before processing the current request [79]. A related model to semi-online framework is the \textit{advice model}, where the \textit{EPI} has been referred to as \textit{bits of advice}. A comprehensive survey on advice models can be found in [92].  

\subsection{Semi-online Scheduling Problem}\label{subsec:Semi-online Scheduling Problem}
Semi-online scheduling [13] is a variant of online scheduling with an \textit{EPI} on future jobs or with additional algorithmic extensions by allowing two parallel policies to operate on each incoming job. It may also include a \textit{buffer} of finite length for pre-processing of a newly arrived job before the actual assignment. We now formally define the semi-online scheduling problem by presenting inputs, constraints and output as follows.
\begin{itemize}
\item Inputs:
\begin{itemize}
\item A sequence $J:<J_1, J_2...,J_n>$ of $n$ jobs with corresponding processing time of $p_i$, where $1\leq i\leq n$ and $p_i> 0$ are revealed one by one for processing on a list $M$=$(M_1, M_2,...,M_m)$ of $m$ parallel machines, where $m\geq 2$ and $n>>>m$.
\item An \textit{EPI} such as arrival order of the jobs or largest processing time or upper and lower bounds on the processing time of the incoming jobs is given a priori.
\end{itemize}
\item Constraints:
\begin{itemize}
\item Each incoming job $J_i$ must be assigned irrevocably to one of the machines $M_j$ as soon as $J_i$ is given.
\item Jobs are non-preemptive, however the preemptive variant of the problem supports job splitting to execute distinct pieces of a job at non-overlapping time spans on the same or different machines.
\end{itemize}
\item Output: Generation of a schedule, representing assignment of all jobs over $m$ machines.(we shall discuss about the output parameters and objectives in section 2.3).
\end{itemize}
\subsection{Practical Applications}\label{subsec: Practical Applications}
Here, we discuss some of the important applications, where semi-online scheduling serves as a major algorithmic framework.
\begin{itemize}
\item \textbf{Resource Management in Operating System} [2]: In a  multi-user, time-shared operating system, it is not known at the outset the sequence of jobs or the number of jobs that would be submitted to the system. Here, jobs are given to the scheduler over time. However, it is the inherent property of the scheduler to make an educated guess about the \textit{maximum and minimum time required to complete a resident job}. The objective is to irrevocably  assign the required computer resources such as memory, processors immediately upon the availability of a job to attain a minimum completion time.     
\item \textbf{Distributed Data Management} [15]: Distributed and parallel systems often confronted to store files of varying sizes on limited capacity remote servers. It is evident that files are submitted from a known source on the fly and each received file must be assigned immediately to one of the remote servers. The central scheduler of the system is handicapped about the successive submissions prior to make an irrevocable assignment. However, it is known for an instance that the submitting source stored the files in $k$ unit capacity servers, which provides a hint for the \textit{total size of files} to be received. The challenge is to store the files on the remote servers with minimum storage requirement.
\item \textbf{Server Request Management or Web Caching} [40]: In a client-server model, it is not known in advance the number of requests that would be submitted to the remote servers nor the time required to process the requests. However, the hierarchical organizations of servers can serve as an extra piece of information for scheduler to cater different level of services to the requests with a broader objective of processing all requests as latest as possible.
\item \textbf{Production and Manufacturing:} Orders from clients arrive on the fly to a production system. The resources such as human beings, machinery equipment(s) and manufacturing unit(s) must be allocated  immediately upon receiving each client order with no knowledge on the future orders. However, one could estimate the \textit{minimum or maximum time} required to complete the order.  Online arrival of the orders have high impact on the renting and purchasing of the high cost machines in the manufacturing units.
\item \textbf{Maintenance and upgrade of industrial tools [52]:} Scheduling of various maintenance and operational activities for modular gas turbine aircraft engines. The goal is to distribute different activities to the machines in such a way that the loads of the the machines will be balanced. The common practice is to maximize load of the least loaded machine.  
\end{itemize}
\subsection{Performance Measure for Semi-online Scheduling}\label{subsec:Performance Measure of Semi-online Scheduling}
Traditional techniques [17] for analyzing the performance of offline scheduling algorithms are largely relied on the entire job sequence, therefore are insignificant in the performance evaluation of semi-online algorithms, which operate on single incoming input at any given time step with minimal knowledge on the future arrivals.\\
\textbf{Competitive analysis} method [8] measures the worst-case performance of a semi-online algorithm $ALG$ designed either for a cost minimization or maximization problem by evaluating \textit{competitive ratio(CR)}. For a cost minimization problem, \textit{CR} is defined as the smallest positive integer $k(\geq 1)$, such that for all valid sequences of inputs in the set $I$= $\{i_1, i_2,...,i_n\}$, we have $C_{ALG}\leq k\cdot C_{OPT}$, where $C_{ALG}$ is the cost obtained by semi-online algorithm $ALG$ for any sequence of $I$ and $C_{OPT}$ is the optimum cost incurred by the optimal offline algorithm $OPT$ for $I$. The \textit{Upper Bound}(UB) on the \textit{CR} obtained by $ALG$ guarantees the maximum value of \textit{CR} for all legal sequences of $I$. The \textit{Lower Bound}(LB) on the \textit{CR} of a \textit{semi-online problem $X$} ensures that there exists an instance of $I$ such that any semi-online algorithm $ALG$ must incur a cost $C_{ALG}\geq b \cdot C_{OPT}$, where $b$ is referred to as \textit{LB} for $X$. The performance of $ALG$ is considered to be \textit{tight}, when $ALG$ ensures no gap between achieved \textit{LB} and \textit{UB} for the problem considered. Sometimes, the performance of $ALG$ is referred to as \textit{tight} if $C_{ALG}$=$k\cdot C_{OPT}$. For a cost maximization problem, \textit{CR} is defined as the infimum $k$ such that for any valid input sequence of $I$, we have $k\cdot C_{ALG}\geq C_{OPT}$.  The objective of a semi-online algorithm is to obtain a \textit{CR} as closer as possible to $1$(strictly greater than or equal to $1$).
\subsection{Research Motivation} 
\label{subsec: Research Motivation}
Research in semi-online scheduling has been pioneered by the following non-trivial issues.
\begin{itemize}
\item The offline $m$-machine($m\geq2$) scheduling problem with makespan minimization objective has been proved to be \textit{NP-complete} by a polynomial time reduction from well-known \textit{Partition} problem [7]. Let us consider an instance of scheduling $n$ jobs on $m$ parallel machines, where $n>>>m$. There are $m^n$ possible assignments of jobs. An optimum schedule can be obtained in worst case with probability $\frac{1}{m^n}$. Further, unavailability of prior information about the whole set of jobs poses a non-trivial challenge in the design of efficient algorithms for semi-online scheduling problems.
\item Given an online scheduling problem considered in the semi-online framework, the non-trivial question raised is:\\
\textit{What can be an additional realistic information on successive jobs that is necessary and sufficient to achieve $1$-competitiveness or to beat the best known bounds on the CR?}
\item A semi-online algorithm is equivalent to an \textit{online algorithm with advice} in the sense that an \textit{EPI} considered in semi-online model can be encoded into bits of advice. The quantification of information into bits will help in analyzing the advice complexity of a semi-online algorithm. Any advancement in semi-online scheduling may lead to significant improvements in the best known bounds obtained by the advice models.
\item Semi-online model is practically significant than the advice model as it considers feasible information on future inputs unlike bits of advice, which sometimes may constitute an unrealistic information.
\end{itemize}

\subsection{ Scope and Uniqueness of Our Survey}
\label{subsec:Scope and Uniqueness of Our Survey }
\textbf{Scope.} This paper surveys scholarly contributions in the design of semi-online scheduling algorithms in various parallel machine models such as identical and uniformly related by considering job's processing formats such as preemptive and non-preemptive with the optimality criteria such as \textit{Min-Max} and \textit{Max-Min}. The aim of the paper is to record important competitive analysis results with the exploration of novel intuitions and critical ideas in a historical chronological  overview.\\
\textbf{Uniqueness.} This is a comprehensive survey article on semi-online scheduling, which describes the motivation towards semi-online scheduling research, outlines a general taxonomy, states fifteen well-known semi-online scheduling algorithms, presents state of the art contributions, explains critical ideas, overviews important results in a chronological manner, organized by \textit{EPI} considered in various semi-online scheduling setups. Several non-trivial research challenges and open problems are explored for future research work. Important references are grouped together in a single article to develop basic understanding, systematic study and to update the literature on semi-online scheduling for future investigation.  
\section{ Taxonomy of Semi-Online Scheduling}\label{sec: Taxonomy of Semi-Online Scheduling}
\begin{figure}[h]
\centering
\includegraphics[scale=0.65]{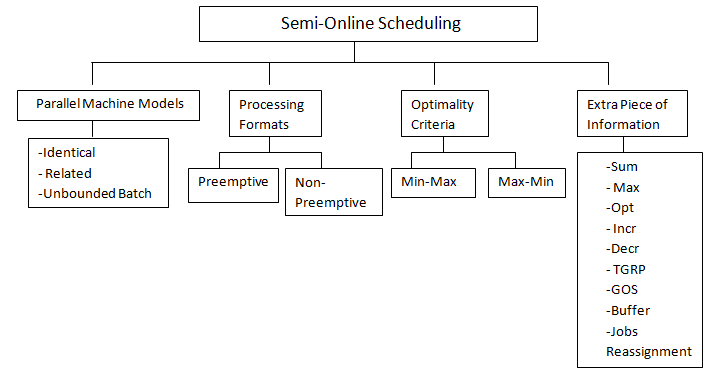}
\caption{A General Taxonomy of Semi-Online Scheduling}
\label{fig:semionlinescheduling.png}  
\end{figure}

The basic terminologies, notations and definitions related to semi-online scheduling are presented in Table \ref{tab:Basic Terminologies Notations and Definition}. 
\begin{table}[]
\caption{Basic Terminologies Notations and Definition}
\begin{tabular} {ccp{8.0cm}}
\hline
\textbf{Terms} & \textbf{Notations} &\textbf{Definitions/Descriptions/Formula} \\
\hline
Job[1] & $J_i$ &  Program under execution, which  consists of a finite number of instructions. A job is also referred to as a collection of at least one smallest indivisible sub task called \textit{thread}. Unless specified explicitly, we assume that a job consists of single thread only. Here, we use terms job and task in the same sense. \\
Machine[1] & $M_j$ & An automated system capable of processing some jobs by following a set of rules. Machine can be a router, web server, robot, industrial tool, processing unit or processor, which is capable of processing the jobs. Here, we use terms machine and processor in the same sense. \\
Processing Time[1-3] &  $p_{ij}$ & Total time of execution of a job $J_i$ on machine $M_j$. For identical machines $p_{ij}$=$p_i$. \\
Largest Processing Time & $p_{max}$ & $\max\{p_i|1\leq i\leq n\}$.\\
Release Time[2, 25] & $r_i$ & The time at which any job $J_i$ becomes available or ready for processing.  \\
Completion Time[3, 25] & $c_i$ & The time at which any job $J_i$  finishes its execution\\
Deadline[3, 25] & $d_i$ & Latest time by which $J_i$ must be finished.\\
Load [11] & $l_j$ & Sum of processing times of the jobs that have been assigned to machine $M_j$. \\
Speed [2, 3] & $S_j$ & The number of instructions processed by  machine $M_j$ in unit time \\
Speed Ratio & $s$ & The ratio between the speeds of two machines. For $2$-machines with speeds  $1$ and $\frac{1}{S}$ respectively. We have speed ratio $s$ = $\frac{1}{\frac{1}{S}} = S$\\
Idle Time [1, 5] & $\varphi$ & The duration of time at which a machine is not processing any task. During the idle time a machine is called idle.\\
Optimal Makespan [2] & $C_{OPT}$ &  $C_{OPT}$= $\max\{p_{max},\hspace*{0.2cm} \frac{1}{m}\cdot \sum_{i=1}^{n}{p_i}\}$\\
\hline
\end{tabular}
\label{tab:Basic Terminologies Notations and Definition}
\end{table} \\
Based on the literature study, a general taxonomy of semi-online scheduling is outlined using the three parameters($\alpha| \beta| \gamma$) based framework of Graham et al. [6] in Figure \ref{fig:semionlinescheduling.png}. Here, $\alpha$ represents  parallel machine models, $\beta$ specifies different job characteristics and $\gamma$ represents optimality criteria. 
\subsection{Parallel Machine Models($\alpha$)}\label{subsec:Machine Models} 
Parallel machine models support simultaneous execution of multiple threads of a single job or a number of jobs on $m$ machines, where $m\geq 2$. Semi-online scheduling problem has been studied in parallel machine models such as identical, uniformly related(or related machines in short) and unbounded batch machines. One model differs from another based on its processing power defined in the literature [6] as follows. 
\begin{itemize} 
\item \textbf{Identical Machines(P)}: Here, all machines have equal speeds of processing any job $J_i$. We have $p_{ij}=p_i$, $\forall M_j, 1\leq j \leq m$. 
\item \textbf{Related Machines (Q)}: Here, the machines operate at different speeds. For a machine $M_j$ with speed $S_j$, execution time of job $J_i$ on $M_j$ is $p_{ij}=\frac{p_i}{S_j}$.
\item \textbf{Unbounded Batch Machine (U-batch)}: A batch machine receives jobs in batches, where a batch($U(t)$) is formed by considering all jobs that are received at time $t$. The jobs in $U(t)$ are processed at the same time in the sense that the completion time($U(c_t)$) of all jobs in a batch are same. The processing time of $U(t)$ is $U(p_t)$=$\max \{p_1, p_2,...,p_k\}$ and the completion time $U(c_t)$=$t+U(p_t)$, where $k$ is the size of $U(t)$ i.e. the number of jobs in a batch. When the size of the batches are not bounded with any positive integers, then it is called \textit{unbounded batch machine} with $k$=$\infty$. 
\end{itemize}
\subsection{Job Characteristics($\beta$)}\label{subsec:Job Characteristics}
\begin{figure}[h]
\centering
\includegraphics[scale=0.55]{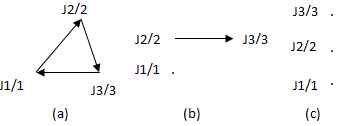}
\caption{(a) Cyclic dependencies among jobs. (b) $J_2$ $\prec$ $J_3$ and $J_1$ is independent of $J_2 $, $J_3$. (c) All jobs are independent.}
\label{fig:precedencegraph.png}  
\end{figure}

Job characteristics describe the nature of the jobs and related characteristics to job scheduling [6, 49]. All jobs of any scheduling problem must possess at least one of the characteristics specified in set $\beta=\{\beta_1, \beta_2, \beta_3, \beta_4, \beta_5\}$. In semi-online scheduling, a new job characteristic $\beta_6$ is  introduced to represent \textit{extra piece of information(EPI)} on future jobs. The job characteristics are presented as follows: $\beta_1$ specifies whether \textit{preemption} or job splitting is allowed, $\beta_2$ specifies precedence relations or dependencies among the jobs, $\beta_3$ specifies release time for each job, $\beta_4$ specifies restrictions related to processing times of the jobs, for example, if $\beta_4$=$1$, that means $ p_{i}=1$, $\forall J_i$,  $\beta_5$ specifies deadline($d_i$) for each job $j_i$, indicating the execution of each $j_i$ must be finished by time $d_i$, otherwise an extra penalty may incur due to deadline over run. Let us throw more clarity on some of the important job characteristics as follows.\\
\textbf{Preemption(pmtn)} allows splitting of a job into pieces, where each piece is executed on the same or different machines in non-overlapping time spans.\\
\textbf{Non-preemption(N-pmtn)} ensures that once a job $J_i$ with processing time $p_i$ begins to execute on machine $M_j$ at time $t$, then $J_i$ continues the execution on $M_j$ until time $t+p_i$ with no interruption in between. \\
\textbf{Precedence Relation} defines dependencies among the jobs by the partial order '$\prec$' rule on the set of jobs [5]. A partial order can be defined on two jobs $J_i$ and $J_k$ as $J_i$ $\prec$ $J_k$, which indicates execution of $J_k$ never starts before the completion of $J_i$. The dependencies among different jobs can be illustrated with a precedence graph $G(p, \prec)$, where each vertex represents a job $J_i$ and labeled with its processing time $p_i$. A directed arc between two vertices in $G(p, <)$ i.e $J_i$ $\rightarrow$ $J_k$ represents $J_i$ $\prec$ $J_k$, where $J_i$ is referred to as \textit{predecessor} of $J_k$. If there exists a cycle in the precedence graph, then scheduling is not possible for the jobs. When there is no precedence relation defined on the jobs, then they are said to be \textit{independent}.
We represent precedence relation among the jobs through precedence graphs by considering three jobs $J_1$, $J_2$, $J_3$ and their dependency relations as shown in Figure \ref{fig:precedencegraph.png}.
\\
\textbf{Extra Piece of Information(EPI)} is the additional information given to an online scheduling algorithm about the future jobs. Motivated by the interactive applications, a number of \textit{EPI}s have been considered in the literature(see the recent surveys [92], [108]) to gain a significant performance improvement over the pure online scheduling policies [92]. 
We now present the definitions and notations of some well studied \textit{EPI}s as follows.
\begin{itemize}
\item \textbf{Sum(\textit{T}).} $\sum_{i=1}^{n}{p_i}$ Total size of all jobs [13].
\item \textbf{Max($p_{max}$).}  $\max \{p_i|1\leq i\leq n\}$ Largest processing time or largest size job [20]. 
\item \textbf{Optimum Makespan(\textit{Opt}).} Value of the optimum makespan is often represented by the following two general bounds [15].\\
\hspace*{2.1cm} $C_{OPT}\geq \frac{1}{m}\cdot\sum_{i=1}^{n}{p_{ij}}$ \hspace*{0.3cm} and \\
\hspace*{2.1cm} $C_{OPT}\geq p_{max}$.
\item \textbf{
Tightly Grouped Processing time (\textit{TGRP}).} Lower and upper bounds on the processing times of all jobs [20].  Some authors [22, 31, 45] considered either lower bound \textit{TGRP(lb)} or upper bound \textit{TGRP(ub}) on the processing times of the jobs.
\item \textbf{Arrival Order of Jobs.} $p_{i+1} \leq p_i$, for $1\leq i\leq n$ Jobs arrive, in order of non-incresing sizes (\textit{Decr}) [21] or $r_{i+1} \geq r_i$  in order of non-decreasing release times (\textit{Incr-r}) [70,87].
\item \textbf{Buffer(B(k)).} A buffer (B(k)) is a storage unit of finite length  $k$($\geq 1$), capable of storing at most $k$ jobs [13]. The weight of B(k) is $w(B(k))\leq \sum_ {i=1}^{k}{p_i}$.  Availability of buffer allows an online scheduling algorithm either to keep an incoming job temporarily in the buffer or to irrevocably assign a job to a machine in case the buffer is full [13]. Therefore, information about $k+1$ future jobs is always known prior to make an efficient scheduling decision. The following variations in the buffer length and usage of buffer have been explored in the literature: buffer with length $k$($\geq 1$) i.e \textit{B(k)} [13], buffer with length 1 i.e \textit{B(1)} [13, 14] and re-ordering of buffer presented as \textit{re B(k)} [56].
\item \textbf{Information on Last Job.} It is known in advance that last job has the largest processing time i.e. $p_n$=$p_{max}$, this \textit{EPI} is denoted by \textit{LL} in [26]. In [28], it is considered that several jobs arrive at the same as last job and this \textit{EPI} is denoted by \textit{Sugg}.
\item \textbf{Inexact Partial Information.} Inexact partial information is also referred to as \textit{disturbed partial information}, which deals with the scenario, where the extra piece of information available to the online algorithm is not exact. For example, the algorithm knows a nearest value of the actual \textit{Sum} but not the exact value. This \textit{EPI} is represented as \textit{$dis Sum$} in [53].
\item \textbf{Reassignment of Jobs($reasgn$).} Once all jobs are assigned to the machines, again they can be reallocated to different machines with some pre-defined conditions. Several conditions on reassignment policies have been proposed in the literature [57, 62] such as reassign the last $k$ jobs, we represent as $reasgn(last(k))$, reassign arbitrary $k$ jobs i.e. $reasgn(k^*)$, reassign only the last job of all machines i.e. $reasgn{(last)}^*$, reassign last job of any one of the machines, represented by $reasgn({(last)}^1)$. 
\item \textbf{Machine availability ($mchavl$).} All machines may not available initially. Machines are available on demand and the release time ($r_j$) of machine $M_j$ is known in advance [64]. Some authors have also considered the scenario where one machine is available for all jobs and other machine is available for few designated jobs [82]. 
\item \textbf{Grade of Service (GOS) or Machine Hierarchy.}  It is known a priori that machines are arranged in a hierarchical fashion to cater different levels of services to the jobs with some defined \textit{GOS} [36, 46]. For example, if a \textit{GOS} of $g_2$ is defined for  any job $J_i$, then $J_i$ can only be assigned to machine $M_2$ and if $J_i$ has \textit{GOS} of $g_1$, then it can be scheduled on any of the machines.    
\end{itemize}
\subsection{Optimality Criteria($\gamma$)}\label{subsec:Optimality Criteria}
\begin{figure}[h]
\centering
\includegraphics[scale=0.55]{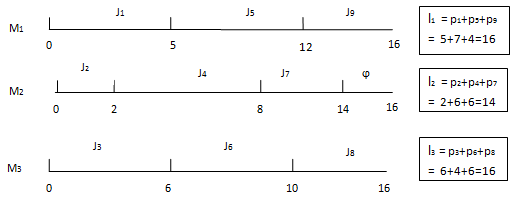}
\caption{Timing Diagram of a Sample Schedule Showing Loads of Machines}
\label{fig:loadprofiles.png}  
\end{figure}

Several optimality criteria or output parameters have been investigated in the offline and online settings [17, 49]. However, in semi-online scheduling the following output parameters have been considered mostly: \textit{makespan} and \textit{load balancing}. 
\begin{itemize} 
\item \textbf{Makespan($C_{max}$)} represents completion time($c_i$) of the job that finishes last in the schedule, $C_{max}$=$\max \{c_i|1\leq i\leq n\}$ or $C_{max}$=$\min\{l_j|1\leq j\leq m\}$. The objective is to \textit{minimize $C_{max}$}, otherwise termed as minimization of the load of highest loaded machine\textit{(min-max)}. 
\item \textbf{Load Balancing} describes the objective to \textit{maximize the minimum machine load(max-min)} or \textit{machine cover}. The scheduler assigns certain number of jobs to each machine for the processing of $n$ jobs on $m$ machines. Each job $J_i$ adds $p_i$ amount of \textit{load} to the assigned machine $M_j$. The goal is to maximize the minimum load occurs on any of the machines so as to keep a balance in the incurred loads among all machines. We refer $C_{min}$ to represent \textit{max-min} objective.  As an example, Figure \ref{fig:loadprofiles.png} shows the loads of machines during the processing of a specified number of jobs. 
\end{itemize}
\textbf{\textit{Examples:}} We present various semi-online scheduling setups based on the three fields ($\alpha| \beta| \gamma$) classification format as shown in Table \ref{tab: Representation of Semi-online Problems }.\\
\begin{table}[!htbp]
\caption {A Sample Format For Representing Semi-online Scheduling Setups}
\begin{tabular}{cp{8cm}}
\hline
\textbf{Setup($\alpha| \beta| \gamma$)} & \textbf{Descriptions} \\
\hline
$P_2| Sum | C_{max}$ &   2-identical machines $|$  no preemption, total processing time  $|$  min-max. \\
$P_2 | Sum, Max | C_{max}$ &  2-identical machines $|$  no preemption, total processing time and maximum size job  $|$ min-max. \\
$P_m | B(k) | C_{max}$ & $m$-identical machines $|$ no preemption, given a buffer of length $k$ $|$ min-max \\
$Q_2| Decr | C_{max}$ & 2-uniform related machines $|$ no preemption, jobs arrive in non-increasing order of their processing times  $|$ min-max. \\
$Q_2| pmtn, TGRP | C_{min}$ & 2-related machine $|$ preemption, lower and upper bounds on the processing times of the jobs $|$ max-min. \\
$U-batch| Max | C_{max}$ & Unbounded batch machine $|$ no preemption, maximum size job $|$ min-max. \\
\hline
\end{tabular}
\label{tab: Representation of Semi-online Problems }
\end{table}\\
\section{Well-Known Semi-Online Scheduling Algorithms}\label{sec:Well-Known Semi-Online Scheduling Algorithms}
For developing a basic understanding on semi-online policies, we represent fifteen well-known semi-online scheduling algorithms as follows. 
\begin{itemize}
\item Algorithm \textbf{$H_1$} was proposed by  Kellerer et al. [13] for $P_2|B(k)|C_{max}$. Algorithm $H_1$ assigns first $k$ incoming jobs to the buffer $B(k)$, where $k \geq 1$. When $({k+1})^{th}$ job arrives, then a job $J_i$ is selected from the buffer, where $J_i \in \{J_1, J_2,.......J_k, J_{k+1}\}$ and is scheduled on machine $M_1$ such that $l_1 + p_i \leq \frac{2}{3} (l_1 + l_2 + w(B))$. If such a $J_i$ does not exist, then any arbitrary job is picked up from the buffer and is assigned to machine $M_2$. Here, $l_1, l_2$ are the loads of machines $M_1, M_2$ respectively before assigning $J_i$ and $w(B)= \sum_{i=1}^{k+1}{p_i}$. 
\item \textbf{Algorithm $H_3$} was proposed by Kellerer et al. [13] for $P_2|Sum|C_{max}$. Algorithm \textbf{$H_3$} schedules each available job $J_i$ on machine $M_1$ as long as $l_1+p_i\leq (\frac{1}{3})\cdot T$, where $T$=$\sum_{i=1}^{n}{p_i}$ and $l_1$ is the load of $M_1$ before the assignment of $J_i$. If $l_1+p_i\leq (\frac{2}{3})\cdot T$, then algorithm $H_3$ schedules $J_i$ on $M_1$ and the remaining jobs $J_t$ are scheduled on machine $M_2$, where $(i+1)\leq t\leq n$.
\item Algorithm \textbf{Premeditated List Scheduling(PLS)} is due to He and Zhang [20] for $P_2|Max|C_{max}$. Algorithm \textit{PLS} assigns each incoming job $J_i$ to machine $M_1$ as long as $p_i\neq p_{max}$ and $l_1 + p_i \leq 2\cdot (p_{max})$, otherwise, $J_i$ is scheduled on machine $M_2$. Thereafter, each incoming job is scheduled on machine $M_j$ for which $l_j$=$\min\{l_1, l_2\}$.
\item \textbf{Algorithm $H$} is due to Angelelli [22] for $P_2|Sum|C_{max}$. Algorithm $H$ assigns an incoming job $J_i$ to machine $M_j$ for which $l_j$=$\max\{l_1, l_2\}$ if $V\geq \max\{|l_1-l_2|, p_i\}$, else schedules $J_i$ on $M_j$ for which $l_j$=$\min\{l_1, l_2\}$, where $V$=$T-(l_1+l_2+p_i)$.
\item Algorithm \textbf{Ordinal} is due to Tan and He [23] for the setup $Q_2|Decr|C_{max}$. Algorithm \textit{Ordinal} schedules all jobs on machine $M_2$ for speed ratio $s\geq (1+\sqrt{3})$. For $s\in [s(k-1), s(k))$, $k\geq 1$, the sub set of jobs $\{J_{ki}, J_{3i+3}|i\geq 0\}$ is scheduled on machine $M_1$ and the sub set of jobs $\{J_1\}\cup \{J_{ki+3}, J_{ki+4},...,J_{ki+k+1}|i\geq 0\}$ is scheduled on machine $M_2$, where $s(k)$=$1$ for $k=0$; $s(k)$=$\frac{1+\sqrt{3}}{4}$ for $k$=$1$ and $s(k)$=$\frac{k^2-1+\sqrt{(k^2-1)^{2}+2k^3(k+1)}}{k(k+1)}$ for $k\geq 2$.
\item Algorithm \textbf{Highest Loaded Machine(HLM)} was proposed by Angelelli et al. [35] and was originally named as \textit{algorithm $H$} for the setup $P_m|Sum|C_{max}$. By observing the behavior of the algorithm, we rename it to \textit{HLM}. Algorithm \textit{HLM} schedules a newly arrive job either on the highest loaded machine in the set of heavily loaded machines or on the highest loaded machine in the set of lightly loaded machine.
\item Algorithm \textbf{Extended Longest Processing Time(ELPT)} was proposed by Epstein and Favrholdt [42] for the setup $Q_2|Decr|C_{max}$. Algorithm \textit{ELPT} assigns each incoming job $J_i$ to the fastest machine $M_j\in \{M_1, M_2\}$ for which $l_j + \frac{p_i}{S_j}$ is minimum, where $S_1$=$1$ and $S_2$=$\frac{1}{s}$ for $s\geq 1$.
\item Algorithm \textbf{Slow LPT} was proposed by Epstein and Favrholdt [42] for $Q_2|Decr|C_{max}$. It schedules the first available job $J_1$ to the slowest machine $M_2$ and the next job $J_2$ is scheduled on the fastest machine $M_1$. It assigns the next incoming job $J_3$ to $M_2$ if $s\cdot (p_1 + p_3) \leq c(s)\cdot (p_2 + p_3)$, otherwise $J_3$ is assigned to machine $M_1$. Next incoming jobs are assigned to the machine $M_j$ for which $l_j + \frac{p_i}{S_j}$ is minimum, where $S_1$=$1$ and $S_2$=$\frac{1}{s}$ for $s\geq 1$. ($c(s)$ is a function of the speed ratio interval $s$) 
\item Algorithm \textbf{Grade of Service Eligibility(GSE)} is due to Park et al. [46] for the setup $P_2|GOS,Sum|C_{max}$. It states that upon the arrival of any job $J_i$ with $GOS=1$, assign $J_i$ to machine $M_1$. When $J_i$ arrives with $GOS=2$, then $J_i$ is assigned to machine $M_2$ if and only if $l_2 + p_i \leq \frac{3}{2}\cdot T$, otherwise, $J_i$ is scheduled on machine $M_1$.
\item Algorithm \textbf{Fastest Last(FL)} was proposed by Epstein and Ye [51] for $P_2|LL|C_{min}$. Algorithm \textit{FL} schedules an incoming job $J_i$ on the slowest machine $M_1$ if and only if $l_2 + p_i > \alpha(S) (l_1 + S\cdot p_i)$, else $J_i$ is scheduled on the fastest machine $M_2$. If $J_i$ is the last job, then it is scheduled exclusively on the fastest machine $M_2$. ($\alpha(S)$ is a function of $S$ and $0 < \alpha(S) < \frac{1}{S}$)
\item Algorithm \textbf{Fractional Semi-online Assignment(FSA)} was proposed by Chassid and Epstein [59] for $Q_2|pmtn,GOS,Sum|C_{min}$. Algorithm \textit{FSA} assigns a newly arrive job $J_i$ with $g_i$=$1$ to machine $M_1$. If $J_i$ has $g_i = 2$, then if $l_2$=$\frac{1}{b+1}$, then $J_i$ is scheduled on $M_1$; else if $l_2+p_i\leq \frac{1}{b+1}$, then $J_i$ is assigned to machine $M_2$; else $\frac{1}{b+1}-l_2$ portion of $J_i$ is assigned to machine $M_2$ and the remaining part of $J_i$ is scheduled on $M_1$. (Note that: $S_1$=$1$, $S_2$=$b$ and $Sum$=$1$, where $b\geq 1$)
\item Algorithm \textbf{RatioStretch} was developed by Ebenlendr and Sgall [61] for $Q_m|pmtn,Decr|C_{max}$. Algorithm \textit{RatioStretch} first estmates for each incoming job $J_i$ the completion time $c_i$=$r\cdot C_{OPT}(i)$, where $r$ is the required approximation ratio and $C_{OPT}(i)$ is the least value of estimated makespan for processing of jobs $J_1, J_2,...,J_i$. Then, two consecutive fastest machines $M_j, M_{j+1}$ are chosen along with time $t_j$ such that if $J_i$ is scheduled on $M_{j+1}$ in the interval $(0, t_j]$ and on $M_j$ from time $t_j$ on wards, then $J_i$ finishes by time $c_i$.
\item Algorithm \textbf{High Speed Machine Priority(HSMP)} was given by Cai and Yang [97] for the setup $Q_2|Max|C_{max}$. Algorithm \textit{HSMP} schedules each incoming job $J_i$ on machine $M_2$ if $p_i$=$p_{max}$; thereafter schedules each incoming $J_{i+1}$ on the machine that will finish $J_{i+1}$ at the earliest; otherwise, schedules $J_i$ on machine $M_1$ if $p_i< p_{max}$ and if $l^{i}_{1}+p_i< l^{i}_{2}+\frac{p_i+p{max}}{s}$, where $l^{j}_{i}$ is the load of machine $M_j$ just before the scheduling of $J_i$ and $1.414\leq s\leq 2.732$; otherwise, schedules $J_i$ on machine $M_2$. 
\item Algorithm \textbf{OM} was proposed by Cao et al. [74] for $P_2|Opt,Max|C_{max}$. It is known at the outset that the first incoming job $J_1$ has the largest processing time $p_{max}$. Algorithm \textit{OM} assigns $J_1$ to machine $M_2$. Thereafter, each incoming job $J_i$, where $2\leq i\leq n$ is scheduled on $M_2$ if and only if $l_2+p_i \leq (\frac{6}{5})\cdot Opt$; otherwise $J_i$ is assigned to machine $M_1$. 
\item Algorithm \textbf{Light Load} was proposed by Albers and Hellwig [75] for $P_m|Sum|C_{max}$. It assigns a new job $J_i$ to the $\lceil \frac{m}{2} \rceil ^{th}$ highest loaded machine $M_j$ if and only if $l_m > 0.25 (\frac{T}{m})$ and $l_j + p_i \leq 1.75 (\frac{T}{m})$; otherwise, $J_i$ is scheduled on the least loaded machine $M_m$.
\end{itemize}       
\section{Historical Overview of Semi-online Scheduling: Important Results}\label{sec:Historical Overview of Semi-online Scheduling: Important Results}
In 1960's, the curiosity to explore computational advantages of multi-processor systems resulted in a number of scheduling models. Online scheduling is one among such models. Graham [2] initiated the study of online scheduling of a list of $n$ jobs on $m(\geq 2)$ identical parallel machines and proposed the famous \textit{List Scheduling(LS)} algorithm. Algorithm \textit{LS} selects the first unscheduled job $J_i$ from the list such that all its \textit{predecessors} ($J_k\prec J_i$) have been completed and schedules $J_i$ on the most lightly loaded machine. Algorithm \textit{LS} achieves performance ratios of $1.5$ for $m$=$2$ and $2-\frac{1}{m}$ for all $m$ by considering $C_{OPT}\geq \frac{\sum_{i=1}^{n}{p_i}}{m}$. In [3], Graham considered the offline setting of $m$-machine scheduling problem and proposed the seminal algorithm \textit{Largest Processing Time(LPT)}. Algorithm \textit{LPT} first sorts the jobs in the list by non-increasing \textit{sizes} and assigns them one by one to a machine that incurs smallest load after each assignment. Algorithm \textit{LPT} achieves a worst-case performance ratio of $1.16$ for $m=2$ and $1.33-\frac{1}{3m}$ for $m\geq 2$ with the time complexity of $O(n\log n)$. These two seminal contributions of Graham served as a motivation for further investigations to address research challenges in online scheduling. Initial three decades(1966-1996) of the online scheduling research were concentrated on the improvements of the \textit{LB} and \textit{UB} on the \textit{CR} to achieve optimal competitiveness(please, see [16-17] for a comprehensive survey on the seminal contributions).  However, no significant attention has been paid for exploring the practicability of the online scheduling model of Graham.\\ 
Motivated by real world applications, Kellerer et al. [13] proposed a novel variant of the online scheduling model by considering  \textit{EPIs} for pre-processing of online arriving jobs and named the variant as semi-online scheduling. They conjectured that additional information on future jobs would immensely help in improving the best competitive bounds in various online scheduling setups. Following the conjecture of Keller et al., ocean of literature have been produced since last two decades in pursuance of achieving optimum competitiveness with the exploration of practically significant new \textit{EPIs}. We now survey the critical ideas and important results given for semi-online scheduling in a historical chronological manner by classifying the results based on the \textit{EPI} as follows.    
\subsection{Early Works in Semi-online Scheduling (1997-2000)} \label{subsec:Early Works on Semi-online Scheduling(1997-2000)}
\textbf{Buffer, Sum.} Kellerer et al. [13] envisioned semi-online scheduling as a theoretically significant and practically well performed online scheduling model. They initiated the study on semi-online scheduling by considering \textit{Sum} as the known \textit{EPI} and proposed algorithm $H_3$ for the setup $P_2|Sum|C_{max}$. Algorithm $H_3$ outperforms algorithm \textit{LS} and achieves a \textit{tight} bound $1.33$ for $m=2$. To show the \textit{LB} $1.33$ of algorithm $H_3$, let us consider an instance of $P_2|Sum|C_{max}$, where $Sum$=$2$. Algorithm $H_3$ schedules each incoming job $J_i$ to machine $M_1$ until $l_1+p_i\leq \frac{2}{3}\cdot$(\textit{Sum}) and assigns the remaining jobs to machine $M_2$. If we consider $Sum$=$2$, then irrespective of the input instances, the final loads $l_1$, $l_2$ would be $\frac{4}{3}$, $\frac{2}{3}$ respectively and $C_{OPT}$ would be $1$. This implies, $C_{H_3} \geq 1.33\cdot (C_{OPT})$. The semi-online strategy of Kellerer et al. unveils that advance knowledge of \textit{Sum} helps any online algorithm $A$ to schedule incoming jobs to a particular machine until its load reaches upto a judiciously chosen fraction of the \textit{Sum} and assigns the remaining jobs to the other machine such that the ratio between $C_A$ and $C_{OPT}$ results in the improved competitive bound. They also studied semi-online scheduling with a buffer($B$) of length $k$ and proposed algorithm $H_1$ for the setup $P_2|B(k)|C_{max}$. They proved that any online scheduling algorithm with $B(k)$ achieves a \textit{CR} of at least $1.33$. The \textit{LB} can be shown by considering an online sequence $J:<J_1/1, J_2/1, J_3/1, J_4/3>$ of four jobs with specified processing times and $k$=$1$. Algorithm $H_1$ keeps the first available job $J_1$ in the buffer. Thereafter, each incoming $J_{i+1}$, where $1\leq i\leq 3$, is either kept in the buffer or any $J_x\in \{J_i, J_{i+1}\}$ is scheduled on machine $M_1$ if $l_1+p_x\leq \frac{2}{3}\cdot (l_1+l_2+w(B))$, else any $J_x$ is assigned to machine $M_2$ (Note that $w(B)$=$p_i+p_{i+1}$). We now have a schedule due to algorithm $H_1$ with the sequence of assignments of $J$ on the machines as follows: $J_1/1$ on $M_1$, $J_2/1$ on $M_2$, $J_4/3$ on $M_1$ and $J_3/1$ on $M_2$ such that $C_{H_1}\geq 4$, where $C_{OPT}\geq 3$. Therefore, $C_{H_1}\geq 1.33\cdot (C_{OPT})$. A matching \textit{UB} was shown to achieve a \textit{tight} bound $1.33$ for algorithm $H_1$. They  also studied a semi-online variant, where two parallel processors are given to virtually schedule a sequence of incoming jobs over $2$-identical machines by two distinct procedures independently. Finally, the jobs are scheduled by the procedure that has incurred minimum $C_{max}$ for the entire job sequence. They obtained a \textit{tight} bound $1.33$ for the semi-online variant $P_2|2$-$Proc|C_{max}$.
Zhang [14] studied the setup $P_2|B(1)|C_{max}$ and obtained the \textit{tight} bound $1.33$ with an alternate policy. The policy keeps the first job $J_1$ in the buffer and if no further jobs arrive, then $J_1$ is scheduled on machine $M_2$, else for next incoming job $J_{i+1}$, where $1\leq i\leq n-1$, the job $J_x$ is chosen from $\{J_i, J_{i+1}\}$ such that $p_x$ is minimum (let us denote the other job as $J_y$). Now, $J_x$ is assigned to machine $M_1$ if $l_1+p_x\leq 2\cdot(l_2+p_y)$, else $J_x$ is scheduled on machine $M_2$. If there is no jobs to arrive further, then the last job in the buffer is assigned to machine $M_2$. The aim of the policy is to keep a larger load difference between machines $M_1$ and $M_2$ by assigning smaller jobs to $M_2$ such that at any time step, the availability and assignment of an unexpected larger job would not incur a makespan beyond $1.33\cdot (C_{OPT})$. Angelelli [22] proposed an alternative to algorithm $H_3$ [13] as \textit{algorithm H} for $P_2|Sum|C_{max}$ and obtained \textit{tight} bound $1.33$. He analyzed the performance of \textit{algorithm H} by considering various ranges of lower bounds on the processing times of the jobs. Girlich et al. [18] obtained an \textit{UB} $1.66$ for $P_m|Sum|C_{max}$. \\
\textbf{TGRP, Max.} He and Zhang [20] initiated study on the setup $P_2|TGRP|C_{max}$ by assuming that all jobs have processing times within the interval of $p$ and $rp$, where $p>0$ and $r\geq 1$. They proved that any online algorithm $A$ must have $C_A\geq \frac{r+1}{2}$ for $r\leq 2$ and $C_A\geq 1.5$ for $r>2$. They analyzed algorithm \textit{LS} for $P_m|TGRP|C_{max}$ and showed that $C_{LS}\leq (1+\frac{(m-1)(r-1)}{m})\cdot C_{OPT}$. They obtained \textit{LB} $1.33$ for the setup $P_2|Max|C_{max}$ by considering the online availability of the job sequence $J:<J_1/1, J_2/1, J_3/2, J_4/2>$, where $p_{max}$=$2$ is known a priori. Following the optimum policy [2] of keeping a machine free for the largest job and assigning the sequence of comparatively shorter jobs to the remaining machines, any online algorithm $A$ assigns $J_1/1$ and $J_2/1$ on $M_1$ followed by the assignments of $J_3/2$ on $M_2$ and $J_4/2$ on either $M_1$ or $M_2$ to incur $C_A\geq 4$, where $C_{OPT}\geq 3$. This implies $C_A\geq 1.33\cdot (C_{OPT})$. They proposed algorithm \textit{PLS} to achieve a \textit{tight} bound $1.33$. Algorithm \textit{PLS} always maintains a load difference maximum of upto $p_{max}$ between machines $M_1$ and $M_2$ such that scheduling of the largest job on the smallest loaded machine almost equalizes the loads of both the machines.\\ 
\textbf{Opt.} Azar and Regev [15] introduced a variant of the classical bin-stretching problem, where items are available one by one in order and each available item must be packed into one of the $m$ bins before the availability of the next item. It is known apriori that all items can be placed into $m$ unit sized bins. The goal is to stretch the bins as minimum as possible so as to fit all items into the bins. Therefore, the bin stretching problem considered by Azar and Regev is analogous to the setup $P_m|Opt|C_{max}$. They achieved an \textit{UB} $1.625$ for large $m$. The key idea is to first define the threshold values $\alpha$ and $2\alpha-1$ based on the value of known \textit{Opt}. Then, arrange $m$ machines in at least two distinct sets based on their loads with respect to $\alpha$ and $2\alpha-1$. A new job is now assigned to the selected machine belongs to the chosen set. Improved rules can be defined for the selections of set and machine. Here, a non-trivial challenge is to define and characterize the threshold values.\\
\textbf{Decr and Preemptive Semi-online Scheduling.} Seiden et al. [21] analyzed algorithm \textit{LPT} [3] with known \textit{Decr} and achieved a \textit{tight} bound $1.166$ for $m$=$2$ and \textit{LB} $1.18$ for $m$=$3$. They initiated the study on $P_m|pmtn, Decr|C_{max}$ and obtained \textit{tight} bound $1.366$. They assumed preemption as \textit{job splitting} for scheduling distinct pieces of an incoming job in the non-overlapping time slots. To understand the notion of job splitting, let us consider a sequence of jobs of unit sizes. Suppose, the required \textit{CR} to be obtained is $k$. Now, the initial $m$ incoming jobs are splitted into at most two pieces each such that all pieces of a job execute in distinct time slots and all jobs are finished by time $k$. Let $r$=$\lfloor\frac{m}{k}\rfloor$, $i\leq r$, each incoming job $J_{m+i}$ is splitted and assigned to the first $i$ machines such that each machine gets $\frac{k}{m}$ fraction of the processing time  of job $J_{m+i}$ and its remaining fraction is assigned prior to time $k$. We now have $i$ highest loads of the machines represented as: $k(1+\frac{i}{m})$, $k(1+\frac{i-1}{m})$,...,$k(1+\frac{1}{m})$, which ensures non-overlapping time slots in the subsequent rounds. Similarly, the next jobs followed by the $({m+r})^{th}$ job are scheduled only in the time slots after $k$ on at most $r+1$ machines. A non-trivial challenge is to rightly choose the values of $r$ and $k$ such that the load to be scheduled prior to time $k$ is at most $k\cdot m$. The authors conjectured that the achieved \textit{tight} bound $1.366$ with known \textit{Decr} can possibly be achieved with only known $p_{max}$. Further, they showed that randomization in scheduling decision making does not lead to improved the \textit{CR} for the setup $P_m|pmtn, Decr|C_{max}$. We now present the main results obtained for semi-online scheduling in identical machines for the years 1997-2000 in table \ref{tab: Important Results: 1997-2000}.    
\begin{table}[h]
\caption {Main Results for Identical Machines: 1997-2000}
\begin{tabular}{|c|p{3.7cm}|p{4cm}|}
\hline
\textbf{Author(s), Year} & \textbf{Setup($\alpha| \beta| \gamma$)}  & \textbf{Competitiveness Results} \\
\hline
Kellerer et al. 1997 [13] & $P_2 | Sum | C_{max}$    $P_2 | B(k) | C_{max}$ $P_2|2-Proc|C_{max}$ & $1.33$ Tight for each setup  \\
\hline
Zhang 1997 [14] &  $P_2 | B(1) | C_{max}$  &  $1.33$ Tight \\
\hline 
Azar and Regev 1998 [15] &  $P_m | Opt | C_{max}$  &  $1.625$ UB \\
\hline
Girlich  et al. 1998 [18] & $P_m | Sum | C_{max}$ & $1.66$ UB \\
\hline
He and Zhang 1999 [20] &   $P_2 | Max | C_{max}$  $P_2 | TGRP[r, rp] | C_{max}$ $P_m|TGRP[r, rp]|C_{max}$   &  $1.33$ Tight with Max, $1.5$ LB for $P_2$ and $r>2$ with TGRP, ($1+\frac{(m-1)(r-1)}{m}$) UB for $P_m$ with TGRP.  \\
\hline
Seiden et al. 2000 [21] &  $P_m | pmtn, Decr | C_{max}$  $P_{2,3} | Decr | C_{max}$  &  $1.366$ Tight for $m \rightarrow \infty$,  $1.166$ Tight for $m=2$, $1.18$ LB for $m=3$  \\ 
\hline
Angelelli 2000 [22] &   $P_2 | Sum,  TGRP(lb) | C_{max}$ & $1.33$ Tight  \\ 
\hline 
\end{tabular}
\label{tab: Important Results: 1997-2000}
\end{table}
\subsection{Well-known Results in Semi-online Scheduling (2001-2005)}
During the years 2001-2005, semi-online scheduling was studied not only for identical machines but for uniform related machines as well. Both preemptive and non-preemptive processing formats were investigated. The concept of combined \textit{EPI} and a new \textit{EPI} on the last job were introduced. We present the state of the art contributions in semi-online scheduling for related machines and identical machines as follows.\\\\
\textbf{Related Machines:}\\
\textbf{Decr.} Tan and He [23] proposed algorithm \textit{Ordinal} for non-preemptive semi-online scheduling with \textit{ordinal data} [11] and known \textit{Decr} for $2$-uniformly related machines, where $S_1$=$1$ and $S_2\geq 1$. They analyzed and proved competitiveness of the algorithm as an interval wise function of machines' speed ratio $s$. They proved the \textit{tightness} of the algorithm in most of the intervals of $s \in [1, \infty)$.  As a main result, they produced \textit{UB} $\frac{s+1}{s}$ for $s\geq 2.732$ and \textit{LB} $\frac{s+1}{s}$ for $s\in [2.732, \infty)$. However, the \textit{LB} of algorithm \textit{Ordinal} does not match with its \textit{UB} when the total length of the speed ratio interval reduces to $0.7784$, where the largest gap between the intervals is at most $0.0521$. Epstein and Favrholdt [30] initiated study on the setup $Q_2|pmtn, Decr|C_{max}$ and achieved competitive ratios of $\frac{3(s+1)}{3s + 2}$ for $ 1 \leq s \leq 3$ and $\frac{2s + (s+1)}{2s^2 + s + 1}$ for $ s \geq 3$. In [42], they investigated the setup $Q_2|Decr|C_{max}$, where $S_1$=$1$ and $S_2$=$\frac{1}{s}$. They expressed competitive ratios as a function of $15$ speed ratio intervals. They proposed algorithm \textit{ELPT} and achieved \textit{tight} bound $1.28$ for $s$=$1.28$. They proposed algorithm \textit{Slow-LPT} for $s\in [1, \frac{1}{6}(1+\sqrt{37})]$ and obtained a \textit{tight} bound $1.28$. Here, the key idea is to initially use the slowest machine and keep the fastest machine free for incoming larger jobs. They proposed algorithms \textit{Balanced-LPT} and \textit{Opposite-LPT} for the remaining intervals, where algorithms \textit{ELPT} and \textit{Slow-LPT} do not obtain tight bounds. Algorithm \textit{Balanced-LPT} schedules the first job $J_1$ on the fastest machine $M_1$. The second job $J_2$ is assigned to machine $M_2$ if $s>c(s)(p_1+p_2)$, else job $J_2$ is scheduled on $M_1$, where $c(s)$=$2.19$ for $s\in [2, 2.19]$ and $c(s)$=$2.57$ for $s\in [2.35, 2.57]$. Thereafter, remaining jobs are scheduled by algorithm \textit{ELPT}. Algorithm \textit{Opposite-LPT} also schedules job $J_1$ on machine $M_1$. The second job $J_2$ is scheduled on $M_1$ if $s\cdot p_2< (p_1+p_2)\leq c(s)\cdot s\cdot p_2$, else $J_2$ is scheduled on $M_2$, where $c(s)$=$2.35$ for $s\in [2.19, 2.35]$. Thereafter, the subsequent jobs are scheduled by the \textit{ELPT} rule. \\
\textbf{Opt.} Epstein [33] studied semi-online scheduling for the setup $Q_2|Opt|C_{max}$, where $S_1$=$1$ and $S_2$=$\frac{1}{s}$. He proposed algorithm \textit{SLOW} by considering $C_{OPT}$=$1$ and $s\geq \sqrt{2}$. Algorithm \textit{SLOW} schedules an incoming job $J_i$ to machine $M_1$ if $l_2\geq \frac{1}{s^2+s}$; else if $l_2+p_j\geq \frac{C_{SL}(s)}{s}$, then job $J_i$ is assigned to machine $M_2$; else $J_i$ is scheduled on machine $M_1$, where $C_{SL}(s)$=$\frac{s+2}{s+1}$. Algorithm \textit{SLOW} performs better in the scenario, where the slowest machine $M_2$ is relatively very slow and initial jobs needs to be assigned to it for keeping the high speed machine $M_1$ relatively free for future larger jobs. For $s\leq \sqrt{2}$, Epstein proposed algorithm \textit{FAST} by considering $C_{OPT}$=$1$. Algorithm \textit{FAST} assigns an incoming job $J_i$ to machine $M_2$ if a job $J_x$ was earlier assigned to $M_2$ due to $l_1<(1+\frac{1}{s}-\frac{C_{FA}(s)}{s})$ and $(l_1+p_x)> C_{FA}(s)$; else if $l_1\geq (1+\frac{1}{s}-\frac{C_{FA}(s)}{s})$, then $J_i$ is scheduled on $M_2$; else if $(l_1+p_j)\leq C_{FA}(s)$, then $J_i$ is scheduled on $M_1$; else $J_i$ is assigned to machine $M_2$, where $C_{FA}(s)$=$\frac{2s+2}{2s+1}$ for $1\leq s\leq \frac{1+\sqrt{17}}{4}$ and $C_{FA}(s)$=$s$ for $\frac{1+\sqrt{17}}{4}\leq s\leq \sqrt{2}$. Algorithm \textit{FAST} performs better in the cases, where the slowest machine $M_2$ is considerably fast, thus allowing initial jobs to be scheduled on the fastest machine $M_1$. He achieved lower bounds in terms of function of defined speed ratio intervals and obtained overall \textit{CR} of $1.414$ and \textit{LB} of $1.366$.\\ 
\textbf{TGRP, Max.} He and Jiang [34] studied the setup $Q_2|pmtn, TGRP(p, xp)|C_{max}$ by considering $S_1$=$1$ and $S_2\geq 1$, where $p>0$ and job size ratio $x\geq 1$. They initiated analysis of algorithm with respect to speed ratio intervals ($s \geq 1$) and job size ratios. They achieved a \textit{tight} bound $\frac{s^2+s}{s^2+1}$ for $s\geq 1$ and $x<2s$. For $s\geq 1$ and $x\geq 2s$, the \textit{tight} bound $\frac{1+2s+s^2}{1+s+s^2}$ was obtained. Further, they investigated the setup $Q_2|pmtn, Max|C_{max}$ by considering known $p_{max}$=$s\geq 1$. They achieved a \textit{CR} of $\frac{2s^2+3s+1}{2s^2+2s+1}$ for $s\geq 1$. They explored that information on \textit{Max} is weaker than known \textit{Decr} for preemptive semi-online scheduling on $2$-related machine. We now present the main results obtained for semi-online scheduling on related machines for the years 2001-2005 in table \ref{tab:Important Results for Uniform Machines: 2001-2005}.
\begin{table}[]
\caption {Main Results for Related Machines: 2001-2005}
\begin{tabular}{|c|p{4.1cm}|p{3.6cm}|}
\hline
\textbf{Author(s), Year} & \textbf{setup($\alpha| \beta| \gamma$)} & \textbf{Competitiveness Results} \\
\hline
Tan and He 2001 [23] & $Q_2 | Decr | C_{max}$ &  $\frac{s+1}{s}$ Tight \\
\hline
Epstein and Favrholdt 2002 [30] & $Q_2 |  pmtn, Decr | C_{max}$ & $\frac{3(s+1)}{3s + 2}$ Tight for $ 1 \leq s \leq 3$, $\frac{2s + (s+1)}{2s^2 + s + 1}$ Tight for $ s \geq 3$   \\
\hline
Epstein 2003 [33] & $Q_2 |  Opt | C_{max}$ & $1.414$ UB, $1.366$ LB \\ 
\hline
He and Jiang 2004 [34] &  $Q_2 | pmtn, Max | C_{max}$  $Q_2 | pmtn, TGRP(p, xp) | C_{max}$  &  $\frac{2s^2 + 3s + 1}{2s^2 + 2s + 1}$ Tight with \textit{Max}, $\frac{s^2 + s}{s^2 + 1}$ Tight with TGRP. \\
\hline
Epstein and Favrholdt 2005 [42] & $Q_2 | Decr | C_{max}$ &  $1.28$ Tight \\
\hline
\end{tabular}
\label{tab:Important Results for Uniform Machines: 2001-2005}
\end{table}\\
\textbf{Identical Machines:}\\
\textbf{Information on Last job(LL).} Zhang and Ye [26] studied a semi-online variant, where it is known apriori that the last job $J_n$ is the largest one i.e. $p_{n}$=$p_{max}$. Upon availability of a job $J_i$, it is also revealed that whether $J_i$ is the last job. They proposed algorithm $A_1$ for the setup $P_2|LL|C_{max}$ and achieved a \textit{tight} bound $1.414$. Algorithm $A_1$ schedules an incoming job $J_i$ on machine $M_2$ if $J_i$=$J_n$; else if $(l_2+p_i)>(0.414)\cdot (l_1+p_i)$, then $J_i$ is assigned to machine $M_1$; else $J_i$ is scheduled on $M_2$. The key idea is to reserve a machine for $J_n$ to obtain relatively minimum makespan irrespective of the size of $J_n$. They proposed algorithm \textit{List Scheduling with a waiting machine(LSw)} for the setup $P_3|LL|C_{max}$ by keeping machine $M_3$ free for $J_n$. Algorithm \textit{LSw} schedules an incoming job $J_i$ on $M_3$ if $J_i$=$J_n$; else assigns job $J_i$ to machine $M_j\in \{M_1, M_2\}$ for which $l_j$=$\min\{l_1, l_2\}$. They proved a \textit{tight} bound $1.5$ for algorithm \textit{LSw}. However, it would be interesting to analyze the cases, where the value of $p_n$=$p_{max}$ is relatively smaller or there are multiple jobs with $p_i$=$p_{max}$.\\
\textbf{Combined Information.} Tan and He [28] exploited the limitation of prior knowledge of $LL$ [26] by considering the following sequences $J:<J_1/1, J_2/1, J_3/2>$ and $J':<J_1/1, J_2/1, J_3/\epsilon>$, where $\epsilon> 0$. They studied the semi-online variants, where two \textit{EPIs} are known at the outset. They proposed the $1.2$ competitive algorithm \textit{SM} for the setup $P_2|Sum, Max|C_{max}$. Algorithm \textit{SM} is designed based on the ratio between known \textit{Sum(T)} and \textit{Max($p_{max}$)}. If $p_{max}\in [\frac{2T}{5}, T]$, then the first job $J_i$ is assigned to machine $M_2$ for which $p_i$=$p_{max}$(such a job is denoted as $J^{1}_{max}$) and other jobs are scheduled on machine $M_1$. If $p_{max}\in (0, \frac{T}{5}]$, then all incoming jobs are scheduled by algorithm \textit{LS}. If $p_{max}\in (\frac{T}{5}, \frac{2T}{5})$, then an incoming job $J_i$ is assigned to machine $M_j\in\{M_1, M_2\}$ such that $(l_j+p_i)\in [\frac{2T}{5}, \frac{3T}{5}]$ and the successive jobs are scheduled on machine $M_{3-j}$. If  $(l_j+p_i)\in [\frac{2T}{5}$-$p_{max}, \frac{3T}{5}$-$p_{max}]$ and if $J^{1}_{max}$ has not been revealed yet, then both $J_i$ and $J^{1}_{max}$ are assigned to $M_j$ and other jobs are scheduled on machine $M_{3-j}$. If $p_i\leq \frac{T}{5}$ or $J_i$=$J^{1}_{max}$, then $J_i$ is scheduled on $M_1$; else if $\frac{T}{5}<p_i\leq p_{max}$, then $J_i$ is scheduled on $M_2$; if two jobs have already been scheduled on $M_2$ such that $l_2\geq \frac{2T}{5}$, then successive jobs are scheduled on $M_1$. Further, they proposed a $1.11$ competitive algorithm for the setup $P_2|Sum, Decr|C_{max}$. They also showed that if \textit{Sum} is given, then information on \textit{LL} is useless and if \textit{Decr} is known, then knowledge of \textit{Max} dos not substantiate to improve the best competitive bound of $1.16$ [21] for $P_2|Decr|C_{max}$. Epstein [33] followed the work of [15] and achieved a \textit{tight} bound $1.11$ for the setup $P_2|Decr, Opt|C_{max}$. He proved the \textit{LB} by considering $C_{OPT}$=$1$ and six jobs, where the jobs $J_1$ and $J_2$ are of \textit{size} $\frac{4}{9}$ each and jobs $J_3, J_4, J_5, J_6$ are of \textit{size} $\frac{5}{18}$ each. If any semi-online algorithm \textit{A} schedules $J_1$ and $J_2$ either on machine $M_1$ or on $M_2$ and schedules the remaining jobs to the other vacant machine, then we have $C_A$=$\frac{10}{9}$. If $J_1$ and $J_2$ are scheduled on two different machines, then by considering the \textit{size} of next three jobs($J'_3, J'_4$, $J'_5$) as $\frac{1}{3}$ each, we have $C_A$=$\frac{10}{9}$ (as any two jobs from $J'_3, J'_4, J'_5$ must be scheduled on a single machine). However, algorithm $OPT$ schedules $J_1$ and $J_2$ on one machine and assigns the remaining three jobs to the other machine to incur $C_{OPT}$=$1$. Therefore, we have $\frac{C_A}{C_{OPT}}$=$\frac{10}{9}$=$1.11$.   He proposed the algorithm \textit{SIZES}, which schedules an incoming job $J_i$ to any $M_j\in \{M_1, M_2\}$ such that $\frac{8}{9}\leq (l_j+p_i)\leq \frac{10}{9}$, the remaining jobs are scheduled on the other machine. If $p_i\leq \frac{2}{9}$, then $J_i$ and all remaining jobs are assigned to machine $M_j$ which incurs $l_j$=$\min\{l_1, l_2\}$ after each assignment.\\
\textbf{List Model.} Yong and Shengyi [27] studied the \textit{list model } of [19], which is a variant of Graham's list scheduling model [2], where it is considered that the machines are not available at the outset. Upon availability of a job $J_i$, an algorithm may purchase a machine by incurring an unit cost. A machine $M_j$ is purchased such that the existing $j$ machines satisfies the following inequality: $l_j\leq T_i<l_{j+1}$, where $T_i$=$\sum_{j=1}^{i}{l_j}$(total work load incurred by initial $i$ jobs). The aim is to optimize the \textit{sum of makespan($C_{max}$) and total machine cost($m$)}. They showed that with known \textit{Max}, an algorithm makes decisions on purchasing of a machine and scheduling of an incoming job $J_i$ by comparing the values of $p_i$, loads of the existing machines or total machine cost with some judiciously chosen bounds on the known $p_{max}$. They obtained an \textit{UB} $1.5309$ and a \textit{LB} $1.33$ with known \textit{Max}. They achieved an \textit{UB} $1.414$ and \textit{LB} $1.161$ with known \textit{Sum}. Further, \textit{List model} can be studied to improve the existing competitive bounds by considering other well-known \textit{EPI}s. We may obtain a natural variant of the \textit{list model} by considering non-identical machines with well defined characteristics, which may influence the choice of an algorithm in purchasing of a machine. \\
\textbf{Max.} Cai [29] extended the work of [20] to obtain a \textit{tight} bound ($\frac{m-2 + \sqrt{(m-2)^2 + 8m^2}}{2m}$) for $P_m|Max|C_{max}$, where $3\leq m\leq 17$. Further, he achieved a \textit{tight} bound $1.414$ for $m\rightarrow \infty$. \\
\textbf{TGRP.} Angelelli et al. [31] considered $TGRP(ub)<1$ and $Sum$=$2$ are known in advance. For $2$-identical machine setup, they obtained lower bounds for various ranges of $ub$. They showed that algorithm \textit{LS} is optimal for smaller $ub$ and proposed optimal algorithms for $0.5\leq ub\leq 0.6$; $ub$=$0.75$ and $0.9\leq ub < 1$. He and Dosa [43] investigated the $3$-identical machine setting by considering $TGRP(p, xp)$, where $p > 0$ and job size ratio $x \geq 1$. They proved that algorithm \textit{LS} is optimal for different intervals of $x \in [1,1.5], [1.73, 2], [6, +\infty]$. They designed algorithms for various ranges of \textit{x} with improved bounds for which the gap between the competitive ratio and the lower bounds is at most $0.01417$. \\
\textbf{Sum, Buffer.} Angelleli et al. [35] extended their previous work [22,31] and obtained an \textit{UB} $1.725$ for $m$-identical machine with known \textit{Sum}. Cheng et al. [44] investigated the setup $p_m|Sum|C_{max}$ by considering \textit{Sum}=$m$. They followed the work of [15, 35] and obtained \textit{UB} $1.6$ and improved \textit{LB} $1.5$ for $m \geq 6$. Dosa et al. [37] studied the setup $P_2|B(1), Sum|C_{max}$ and obtained a \textit{tight} bound $1.25$. They showed that considering a $B(k)$, where $k>1$ does not help to improve the $1.25$ competitiveness. They explored that when \textit{Sum} is known at the outset, then the \textit{knowledge of the sizes of $k(>1)$ future jobs($k$-look ahead)} does not help in improving the competitive bound. Further, they studied the setup $P_2|2$-$Proc|C_{max}$ with known \textit{Sum} and improved the \textit{tight} bound $1.33$ obtained in [13] to $1.2$. We now present the main results obtained for semi-online scheduling on identical machines for the years 2001-2005 in table \ref{tab: Important Results on Identical Machines: 2001-2005}.  
\begin{table}[h]
\caption {Important Results for Identical Machines: 2001-2005}
\begin{tabular}{|c|p{3.6cm}|p{4.5cm}|}
\hline
\textbf{Author(s), Year} & \textbf{setup($\alpha| \beta| \gamma$)} & \textbf{Competitiveness Results} \\
\hline
Zhang and Ye 2002 [26] & $P_2 | LL | C_{max}$  $P_3 | LL | C_{max}$ & $1.414$ Tight  for $m=2$, $1.5$ Tight for $m=3$. \\
\hline 
Yong and shengyi 2002 [27] &  $P_m | Max | C_{max}$+$m$  $P_m | Sum | C_{max}$+$m$ &  $1.53$ UB and $1.33$ LB with \textit{Max},  $1.414$ UB and $1.161$ LB with \textit{Sum}  \\
\hline
Tan and He 2002 [28] &  $P_2 | Sum, Max | C_{max}$ $P_2 | Sum, Decr | C_{max}$ &  $1.2$ Tight with \textit{Sum} and \textit{Max }, $1.11$ Tight with \textit{Sum} and \textit{Decr}\\
\hline 
Cai 2002 [29] & $P_m |  Max | C_{max}$ &  ($\frac{m-2 + \sqrt{(m-2)^2 + 8m^2}}{2m}$) Tight for $3 \leq m \leq 17$, $1.414$ Tight for $m\rightarrow \infty$. \\
\hline 
Angelelli et al. 2003 [31] & $P_2 | Sum, TGRP(ub) | C_{max}$ & $1.2$ Tight for $ub \in [0.5, 0.6]$ , ($1 + (\frac{ub}{3}$)) Tight for $ub \in [0.75, 1]$,   ($0.666(1+ ub)$) Tight for $ub \in [0.94, 1]$ . \\
\hline
Epstein 2003 [33] & $P_2 |  Decr, Opt | C_{max}$ & $1.11$ Tight. \\ 
\hline
Angelelli et al. 2004 [35] &  $P_m | Sum | C_{max}$ &  $1.725$ UB,  $1.565$ LB for $m \rightarrow \infty$ \\
\hline
Dosa et al. 2004 [37] &  $P_2 | B(1), Sum | C_{max}$ $P_2 | 2$-$Proc, Sum | C_{max}$
 &  ($1.25$, $1.2$) Tight for respective setups\\
\hline
He and Dosa 2005 [43] & $P_3 | TGRP | C_{max}$  &  $1.5$ Tight for $x \in (2, 2.5]$, ($\frac{4r +2}{2r + 3}$) Tight for $x \in (2.5, 3]$, ($1.66 - \frac{\delta}{18}$) Tight for $x \in (3, 6)$ \\
\hline
Cheng et al. 2005 [44] &  $P_m |  Sum | C_{max}$ &   ($1.6, 1.5$) UB and LB respectively for $m \geq 6$ \\
\hline 
\end{tabular}
\label{tab: Important Results on Identical Machines: 2001-2005}
\end{table}
\subsection{Advancements in Semi-online Scheduling (2006-2010)} \label{sebsec:Advancements in Semi-online Scheduling(2006-2010)}  The initial decade in  semi-online scheduling research was devoted to the traditional online scheduling setups with fundamental \textit{EPI}s on the future jobs. Moreover, the significance of \textit{EPI} was realized with the improvement in the competitive bounds for pure online scheduling setups. During the years 2006-2010, new semi-online scheduling setups such as \textit{GOS} or machine hierarchy; a variant of \textit{EPI} such as \textit{inexact EPI} and new policies such as \textit{job re-assignment} and \textit{buffer re-ordering} have been introduced. We now discuss on the important results contributed during the years 2006-2010 for semi-online scheduling on identical and related machines as follows. \\
\textbf{Identical Machines:}\\
\textbf{Sum.} Angelelli et al. [45] studied the setup $P_2|Sum, TGRP(ub)|C_{max}$ and advanced their previous work [31] for the unexplored intervals of $ub$. They showed \textit{LB}s for the interval, where $ub\in [\frac{1}{k}, \frac{1}{k-1}]$ and $k\geq 2$. For an instance, a \textit{LB} of $(\frac{k-1}{3})\cdot ub$+$\frac{2}{3}\cdot(\frac{k+1}{k})$ was shown for $ub\in [\frac{2(k+1)}{k(2k+1)}, \frac{2k-1}{2k(k-1)}]$. The \textit{LB} was proved by considering two job sequences $J'$ and $J''$, where $J'$=$\{J_1/x, J_2/x, J_3/y, J_4/y$ and $2(k-1)$ jobs of size $ub \}$, where $x\in [0, ub]$ such that $x+y+(k-1)\cdot ub$=$1$ and $y\leq x\leq \frac{ub}{2}$ and $J''$=$\{J_1/x, J_2/x, J_3/z$ and $2(k-1)$ jobs of size $\frac{1}{k} \}$, where $2x+z+(k-2)\cdot \frac{1}{k}$=$1$ and $x<z<\frac{1}{k}$. Any algorithm \textit{A} has the option to schedule the fist two jobs $J_1$ and $J_2$ either on the same machine or on different machines. If algorithm \textit{A} schedules $J_1$ and $J_2$ on the same machine, then for the sequence $J'$, we obtain $C_A\geq 2x+(k-1)\cdot ub$. If $J_1$ and $J_2$ are scheduled on different machines, then for $J''$ we have $C_A\geq x+(k-1)\cdot \frac{1}{k}+z$=$1-x+\frac{1}{k}$. Therefore, in both cases, we obtain $C_A\geq \min\{1-x+\frac{1}{k}, 2x+(k-1)\cdot ub\}$. We obtain $C_{OPT}$=$1$ by assigning $J_1$ and $J_2$ to different machines for $J'$ and by scheduling them on the same machine for $J''$. Therefore, we have $\frac{C_A}{C_{OPT}}$=$\min\{1-x+\frac{1}{k}, 2x+(k-1)\cdot ub\}$. By maximizing w.r.t $x$, we achieve $\frac{C_A}{C_{OPT}}\geq (\frac{k-1}{3})\cdot ub$+$\frac{2}{3}\cdot(\frac{k+1}{k})$. They proposed the optimal algorithm $H'$ for $ub\in [\frac{1}{k}, \frac{2(k+1)}{k(2k+1)}]$, which is $(k\cdot ub)$-competitive for $ub\in [\frac{2(k+1)}{k(2k+1)}, \frac{1+2k}{2k^2})$. Algorithm $H'$ schedules an incoming job $J_i$ on the machine $M_1$ if $l_1+p_i\leq 1+\frac{1}{2k+1}$; else if $l_2+p_i\leq 1+\frac{1}{2k+1}$, then $J_i$ is scheduled on the machine $M_2$; else $J_i$ is assigned to the machine $M_j\in \{M_1, M_2\}$ such that $l_j$=$\min\{l_1, l_2\}$. They also proposed a $(1+\frac{1}{2k})$-competitive algorithm for $ub\in [\frac{1+2k}{2k^2}, \frac{2k-1}{2k(k-1)}]$. In [50], they studied the setup $P_3|Sum|C_{max}$ and obtained the \textit{LB} ($1+ \frac{\sqrt{129}-9}{6}) > 1.392$. An \textit{UB} $1.421$ was shown by a pre-processing policy of the available jobs. Here, a non-trivial challenge is to tighten or diminish the gap between the obtained \textit{LB} and \textit{UB}. \\
\textbf{Max.} Wu et al. [58] followed the work of [29] and obtained a \textit{tight} bound $2-\frac{1}{m-1}$ for $m$=$3, 4$ with known \textit{Max}. Sun and Huang [64] considered a variant, where all machines are not given at the outset. However, machine availability time $r_j$ is given at the outset for each machine. W.l.o.g, it is assumed that $r_m\geq r_{m-1}\geq....\geq r_1$. They obtained a \textit{LB} $1.457$ for $m>6$. They proposed a $(2-\frac{1}{m-1})$-competitive algorithm, which assigns an incoming job $J_i$ by algorithm \textit{LS} unless $p_i$=$p_{max}$ and $(r_{min}+l_{min}+p_i)> 2\cdot p_{max}$; otherwise $J_i$ is scheduled on machine $M_1$ and the successive jobs are scheduled by algorithm \textit{LS}, where $r_{min}$ and $l_{min}$ are the release time and load of the most lightly loaded machine respectively.\\
\textbf{Combined Information.}  Hua et al. [48] advanced the work of [28] for $3$-identical machine setting with known \textit{Sum} and \textit{Max}. They obtained an \textit{UB} $1.4$ and a \textit{LB} $1.33$. Wu et al. [54] tighten the gap between the obtained \textit{UB} and \textit{LB} of [48] and obtained a \textit{tight} bound $1.33$ for the setup $P_3|Sum, Max|C_{max}$. \\
\textbf{GOS.} Park et al. [46] initiated the study on semi-online scheduling under \textit{GOS} eligibility with known \textit{Sum}. They considered that a job with $g_i$=$1$ can only be processed by machine $M_1$ and if $g_i$=$2$, then $J_i$ can be processed by any of the machines. They proposed a $1.5$-competitive semi-online algorithm for the setup $P_2|GOS, Sum|C_{max}$. The algorithm schedules an incoming job $J_i$ to machine $M_1$ if $g_i$=$1$; else if $g_i$=$2$ and $l_2+p_j\leq (\frac{3}{2})\cdot L$, then $J_i$ is scheduled on machine $M_2$; else $J_i$ is assigned to machine $M_1$, where $L$=$\frac{Sum}{2}$. For the same problem, Jiang et al. [47] studied the preemptive version with \textit{GOS} and proposed a $1.5$ competitive algorithm. For the non-preemptive case with \textit{GOS}, they improved the \textit{UB} from $2$ obtained in [10] to $1.66$. Wu and Yang [66] studied $2$-identical machine case with \textit{GOS}. They investigated the problem separately for known \textit{Opt} and \textit{Max}.\\
\textbf{Inexact EPI.} Tan and He [53] studied semi-online settings, where the value of a known \textit{EPI} is given in interval or in the inexact form unlike the exact value. For some $x>0$ and the \textit{disturbance parameter} $y\geq 1$, the following \textit{EPI}s were considered for the respective settings: for $P_2|disOpt|C_{max}$, it is given that $C_{OPT}\in [x, yx]$; for $P_2|disSum|C_{max}$, it is known that $Sum\in [x, yx]$ and for $P_2|disMax|C_{max}$, it is known that $p_{max}\in [x, yx]$. For $P_2|disOpt|C_{max}$, they achieved a \textit{LB} $1.5$ for $y\geq 1.5$ and obtained \textit{UB}s $\frac{7y+1}{4y+2}$ for $1\leq y\leq \frac{5+\sqrt{41}}{8}$ and $y$ for $\frac{5+\sqrt{41}}{8}\leq y< 1.5$. They proved \textit{LB} $1.5$ for the setup $P_2|disSum|C_{max}$, where $y\geq 1.5$. For $P_2|disMax|C_{max}$, they proved \textit{LB}s $\frac{2y+2}{y+2}$ for $y$=$1.23$ and $1.5$ for $y\geq 2$. Further, they proposed the algorithm \textit{modified PLS(MPLS)} and achieved an \textit{UB} $\frac{2y+2}{y+2}$ for $y\in [1, 2]$ and showed its tightness for $y\in [1, \sqrt{5}-1]$. Algorithm \textit{MPLS} assigns each incoming job $J_i$ to machine $M_1$ until the arrival of any job $J_b$ for which $p_b\in [1, y]$ and $(l_1+p_b)>2$. Thereafter, $J_b$ and all successive jobs are scheduled by algorithm \textit{LS}.  \\
\textbf{Job Reassignment.} Tan and Yu [57] studied a semi-online variant, where an algorithm is allowed to re-schedule some of the already assigned jobs under certain conditions. For the setup $P_2|reasgn(last(k))|C_{max}$, they proved \textit{LB} $1.5$ and showed that algorithm \textit{LS} is optimal with no re-assignments. For $P_2|reasgn(last)|C_{max}$, they proposed algorithm \textit{RE} and obtained a \textit{tight} bound $1.414$. Algorithm \textit{RE} assigns an incoming job $J_i$ to the highest loaded machine $M_j$ if $l_1\leq (\sqrt{2}+1)\cdot (l_2+p_i)$ and $p_i\leq \sqrt{2}\cdot l_1$; otherwise, $J_i$ is scheduled on machine $M_{3-j}$. After the assignment of all jobs, algorithm \textit{RE} checks for re-assignment. If all jobs have been scheduled on the same machine $M_j$, then the job $J_n$(last job) is re-scheduled on the machine $M_{3-j}$. Let $J^{1}_{n_1}$ and $J^{2}_{n_2}$ be the last two jobs scheduled on machines $M_1$ and $M_2$ respectively. Let us consider $p_x$=$\max\{p^{1}_{n_1}, p^{2}_{n_2}\}$ and $p_y$=$\min\{p^{1}_{n_1}, p^{2}_{n_2}\}$. Algorithm \textit{RE} re-assigns $J_x$ followed by $J_y$ to the $M_j$, which can obtain minimum $c_x$ and $c_y$ respectively. For $P_2|reasgn(k^*)|C_{max}$, they proposed algorithm \textit{RA} and achieved a \textit{tight} bound $1.33$. Algorithm \textit{RA} schedules jobs $J_1$ and $J_2$ on two different machines. Let $l_1$=$\max\{l_1, l_2\}$. Each incoming job $J_i$, where $3\leq i\leq n$, is scheduled on machine $M_1$ if $l_1+p_i\leq 2\cdot l_2$; otherwise, $J_i$ is scheduled on machine $M_2$. After the scheduling of all jobs, if $l_2> 2\cdot l_1$, then the job $J^{2}_{n_2-1}$ is re-scheduled on machine $M_1$.
The following non-trivial questions remain open: What is the minimum number of re-assignments that is sufficient to improve the known competitive bounds? Is the re-assignment policy with \textit{EPI} such as \textit{Decr}, \textit{Opt}, \textit{Sum} or \textit{Max} practically significant and helps in achieving optimal bounds on the \textit{CR}? \\
\textbf{Max-Min Objective.} Tan and Wu [52] studied non-preemptive semi-online scheduling on $m$-identical machine($m \geq 3$) with $C_{min}$ objective. They proposed a $(m-1)$-competitive algorithm for the setup $P_m|Sum|C_{min}$. The idea is to keep the loads of all machines under $\frac{Sum}{2m}$. The machine $M_m$ is reserved from  starting to schedule a job $J_i$, if there exists no machine $M_j$, where $1\leq j\leq m-1$ for which $l_j$ is at most $\frac{Sum}{m}$ or $\frac{Sum}{2m}$ after the assignment of $J_i$. If there exists some machines with load at most $\frac{Sum}{m}$, then assignment of $J_i$ to $M_m$ makes $l_m> \frac{Sum}{2m}$ and if there are some machines with load at most $\frac{Sum}{2m}$, then $J_i$ and the remaining jobs are scheduled on $M_m$. They proposed a $(m-1)$-competitive algorithm for $P_m|Max|C_{min}$. Each incoming job is scheduled on any one of the $m-1$ machines by algorithm \textit{LS} until the arrival of a job $J_i$ with $p_i$=$p_{max}$ or $p_i+\min\{l_1, l_2,...,l_{m-1}\}> 2\cdot (p_{max})$. Such a $J_i$ is scheduled on machine $M_m$ and the successive jobs are scheduled over $m$-machines by algorithm \textit{LS}. The idea is to maintain a load of at most $2\cdot (p_{max})$ in each machine, where the machine $M_m$ is kept idle for the largest job $J_b$ with $p_b$=$p_{max}$. They obtained \textit{tight} bounds $1.5$ and $m-2$ for $m$=$3$ and $m\geq 4$ respectively with combined information on \textit{Sum} and \textit{Max}. We now present the main results obtained for semi-online scheduling on identical machines for the years 2006-2010 in table \ref{tab: Important Results on Identical Machines: 2006-2010}. 
\begin{table}[h]
\caption {Important Results for Identical Machines: 2006-2010}
\begin{tabular}{|c|p{3.6cm}|p{4.8cm}|}
\hline
\textbf{Author(s), Year} & \textbf{setup($\alpha| \beta| \gamma$)} & \textbf{Competitiveness Results} \\
\hline
Angelelli et al. 2006 [45] & $P_2 | Sum, TGRP(ub) | C_{max}$ & ($1+ \frac{1}{2n + 1}$) Tight for $ub \in [ \frac{1}{n}, \frac{2(n+1)}{n(2n+1)}]$, ($(\frac{n-1}{3}) ub + (0.666) (\frac{n+1}{n})$) Tight for $ub \in (\frac{2n-1}{2n(n-1)}, \frac{1}{n-1}]$ \\
\hline
Park et al. 2006 [46] & $P_2 |  GOS, Sum | C_{max}$  &   $1.5$ Tight \\
\hline
Jiang et al. 2006 [47] & $P_2 |  GOS | C_{max}$ $P_2 |  pmtn, GOS | C_{max}$ & $1.66$ UB, $1.5$ Tight \\
\hline
Hua et al. 2006 [48] &  $P_3 |  Sum, Max | C_{max}$ & $1.4$ UB, $1.33$ LB. \\
\hline
Angelelli et al. 2007 [50] & $P_3 | Sum | C_{max}$ & $1.392$ LB, $1.421$ UB. \\
\hline
Tan and Wu 2007 [52] & $P_m | Sum | C_{min}$ $P_m | Max | C_{min}$ $P_m | Sum, Max | C_{min}$ & ($m-1$)-competitive for \textit{Sum} or \textit{Max}, $1.5$ Tight with \textit{Sum} and \textit{Max} for $m=3$, ($m-2$) Tight for $m \geq 4$. \\ 
\hline
Tan and He 2007 [53] & $P_2 | dis Opt | C_{max}$ $P_2 | dis Sum | C_{max}$ $P_2 | dis Max | C_{max}$ & $1.5$ Tight with \textit{Opt} or \textit{Sum} for $y \geq 1.5$, $1.5$ Tight with \textit{Max} for $y \geq 2$\\
\hline
Wu et al. 2007 [54] & $P_3 | Sum, Max | C_{max}$ & $1.33$ Tight. \\
\hline
Tan and Yu 2008 [57] & $P_2 | reasgn(last(k)) | C_{max}$ $P_2 | reasgn(k^*) | C_{max}$ $P_2 | reasgn(last) | C_{max}$ & ($1.5$, $1.33$, $1.414$) LB for respective setups \\
\hline
Wu et al. 2008 [58] & $P_m | Max | C_{max}$ & ($2- \frac{1}{m-1}$ ) Tight \\
\hline
Sun and Huang 2010 [64] & $P_m | r_j, Max | C_{max}$ & $1.457$ LB for $m > 6 $,  ($2- \frac{1}{m-1}$) Tight. \\
\hline
Wu and Yang 2010 [66] & $P_2 | GOS, Max | C_{max}$ $P_2 | GOS, Opt | C_{max}$ & ($1.618, 1.5$) Tight for respective setups. \\
\hline
\end{tabular}
\label{tab: Important Results on Identical Machines: 2006-2010}
\end{table} \paragraph{}
\textbf{Related Machines:}\\
\textbf{Last Job.} Epstein and Ye [51] followed the work of [26] and considered \textit{LL} as the known \textit{EPI} in their study of semi-online scheduling on $2$-related machines with \textit{min-max} and \textit{max-min} optimality criteria. They considered $S_1$=$\frac{1}{s}$ and $S_2$=$1$, where $s\geq 1$. They proposed in general an algorithm for both optimality criteria and analyzed its performance for various intervals of $s$. The algorithm schedules an incoming job $J_i$ on machine $M_1$ if $l_2+p_i> \alpha(s)\cdot (l_1+s\cdot p_i)$; otherwise job $J_i$ is scheduled on machine $M_2$, where $0< \alpha(s)< \frac{1}{s}$. If $J_i$=$J_n$, then $J_i$ is scheduled on $M_2$. The key idea is to keep the highest speed machine $M_2$ relatively light loaded to schedule $J_n$(\textit{largest job}) on it. They obtained \textit{tight} bound $2.618$ for the setup $Q_2|LL|C_{min}$. They achieved \textit{UB} $1.5$ and \textit{LB} $1.465$ for the setup $Q_2|LL|C_{max}$. \\
\textbf{Sum.} Angelelli et al. [55] studied the setup $Q_2|Sum|C_{max}$. They considered speeds $S_1$=$x$, $S_2$=$1$ and $Sum$=$1+x$, where $x\geq 1$. They showed \textit{LB} and \textit{UB} as functions of $x$. They proposed algorithm $H'$ for $x\in [1, 1.28]$, which assigns an incoming job $J_i$ to machine $M_1$ if $l_1+p_i\leq x\cdot (1+\frac{1}{2x+1})$; otherwise $J_i$ is scheduled on machine $M_2$. They proved $(\frac{2+2x}{2x+1})$-competitiveness of algorithm $H'$. They developed algorithm $H''$ for $x\in[1.28, 1.41]$, which assigns an incoming job $J_i$ to machine $M_1$ if $l_1+p_i \leq x^2$; else $J_i$ is assigned to machine $M_2$. They showed that algorithm $H''$ is $x$-competitive. For $x\geq 1.41$, they designed algorithm $H'''$, which assigns an incoming job $J_i$ to machine $M_2$ if $l_2+p_i\leq 1+\frac{1}{x+1}$; else $J_i$ is scheduled on machine $M_1$. They proved ($\frac{x+2}{x+1}$)-competitiveness of algorithm $H'''$. Ng et al. [60] studied the setup $Q_2|Sum|C_{max}$ by considering $S_1$=$1$ and $S_2\geq 1$. They obtained competitive bounds as functions of intervals of $s\geq 1$, where the largest gap between the \textit{LB} and \textit{UB} is at most $0.01762$. They achieved a \textit{LB} $\frac{s+2}{s+1}$ for $s\geq \sqrt(3)$ and overall \textit{UB} $1.369$ for $s \in [1, \infty)$.  Angelelli et al. [63] investigated for the setup stated in [55, 60] by introducing a \textit{geometric representation} of the scheduling problem through a \textit{planar model}. They considered $2$-related machine setup with speeds $S_1$=$1$, $S_2$=$b$ and $Sum$=$b+1$, where $b\geq 1$. They represented scheduling of jobs in planar model as a game between \textit{constructor(K)} and \textit{scheduler(H)}, where \textit{K} submits jobs one by one and \textit{H} schedules a job upon its availability on a machine by following an algorithm. They illustrated the game in a plane by representing each point($x, y$) as the situation, where $x$=$l_1$ and $y$=$l_2$. Here, a move of \textit{K} corresponds to the arrival of a new job $J_i$ with $p_i>0$ and the move of \textit{H} specifies, whether to move to the point($x+p_i, y$) or to the point ($x, y+p_i$) from the point ($x, y$) in the plane. The game ends after reaching the line $x+y$=$b+1$. Now, the current position of the point($x, y$) determines the makespan incurred by the scheduler  $H$. They showed a \textit{LB} $1.359$ for $b\in[1.366, 1.732]$, which they proved to be optimal for $b$=$1.5$.\\
\textbf{Buffer.} Englert et al. [56] investigated both $m$-identical and $m$-related machines settings with a buffer of size $k\in \theta(m)$(where, $\theta(m)$ is a function on number of machines). They introduced the \textit{re-ordering of buffer} policy which does not assign each incoming job immediately to any of the machines, rather stores the jobs in the buffer and re-order the stored input job sequence prior to construct the actual schedule so as to achieve minimum makespan. They obtained \textit{LB} and \textit{UB} $1.333$, $1.465$ respectively for $m$-identical machine which beats the previous best results obtained by non-reordering buffer strategies of [13, 14, 37]. For $m$-related machine setup, they obtained an \textit{UB} $2$ with a buffer of size $m$. \\ 
\textbf{Preemptive Semi-online Scheduling.} Chassid and Epstein [59] studied preemptive semi-online scheduling on $2$-related machine setup. They considered both \textit{max-min} and \textit{min-max} optimality criteria with known \textit{GOS} and  \textit{Sum}. They considered that $S_1$=$1$, $S_2$=$b$, $Sum$=$1$, where $b\geq 1$. They assumed that a job $J_i$ with $g_i$=$1$ must be processed only on machine $M_1$ and with $g_i$=$2$ it can be processed on any $M_j\in\{M_1, M_2\}$. They proposed $1$-competitive algorithm \textit{FSA} and proved its \textit{tightness} for both optimality criterion with the key idea of keeping the load of machine $M_2$ at most $\frac{Sum}{b+1}$. The optimality of algorithm \textit{FSA} was shown by analyzing the following two cases. Case 1: if $l_2$=$\frac{Sum}{b+1}$, then we have $l_1$=$\frac{Sum}{b+1}$ by considering $Sum$=$1$ and $b$=$1$, this implies $C_{FSA}$=$\frac{1}{2}$ followed by $\frac{C_{FSA}}{C_{OPT}}$=$1$, where $C_{OPT}\geq \frac{Sum}{2}$=$\frac{1}{2}$. Case 2: if $l_2<l_1$, means machine $M_2$ has been equipped with all $J_i$s' with $g_i$=$2$, so the remaining $J_i$s' with $g_i$=$1$ have been scheduled on machine $M_1$, which eventually balances the loads of $M_1$ and $M_2$. Therefore, we obtain optimal schedules in both the cases. Ebenlendr and sgall [61] proposed an unified algorithm \textit{RatioStretch} for preemptive semi-online scheduling on $m$-uniform machines($m\geq 2$). They proved that the algorithm achieves optimum approximation ratio that holds for any values of $s$ with any known \textit{EPI}. They computed the ratio by linear program, where machines speeds are considered as input parameters. They established relationships among well-known semi-online setups for uniform machines and obtained competitive bounds in each setup for large $m$.\\  
\textbf{Opt.} Ng et al. [60] improved the results of Epstein [33] for the speed ratio interval $s\in [1.366, 1.395]$ and obtained an \textit{UB} $\frac{2s+1}{2s}$. They showed \textit{tight} bound $1.366$ for overall $s\in [1, \infty)$. \\
\textbf{Job Reassignment.} Liu et al. [62] studied the setup $Q_2|GOS|C_{max}$ by considering $S_1$=$1$ for higher \textit{GOS} machine($M_1$) and $S_2$=$x$ for ordinary machine($M_2$). They obtained \textit{LB}s $1+ \frac{2x}{x+2}$ for $0 < x\leq 1$ and $1+ \frac{x+1}{x(2x+1)}$ for $x > 1$ by considering different \textit{GOS} levels. They proved \textit{LB} $1+ \frac{1}{1+x}$ with \textit{re-assignment of last $k$ jobs(reasgn(last(k)))} and \textit{LB} $\frac{{(S+1)}^2}{S^2 + S + 1}$ for \textit{re-assignment of  one job from every machine(reasgn$(last)^*$)}. They proposed ($\frac{{(x+1)}^2}{x+2}$)-competitive algorithm \textit{EX-RA } for both types of re-assignment policies  by considering $S_1$=$x$ and $S_2$=$1$, where $1\leq x \leq 1.414$. Algorithm \textit{EX-RA} schedules the jobs $J_1$ and $J_2$ on  different machines such that $l_1$=$\max\{p_1, p_2\}$ and $l_2$=$\min\{p_1, p_2\}$. For each incoming job $J_i$($3\leq i\leq n$), if $l_j+\frac{p_i}{x}\leq (x+1)\cdot l_2$, then job $J_i$ is scheduled on machine $M_1$; otherwise $J_i$ is assigned to machine $M_2$. After the scheduling of job $J_n$, if $l_2\leq (x+1)\cdot l_1$, then we have $C_{EX-RA}$=$\max\{l_1, l_2\}$; otherwise the second last job of machine $M_2$ is re-scheduled on machine $M_1$ and $l_1$, $l_2$ is updated to obtain the final $C_{EX-RA}$=$\max\{l_1, l_2\}$. Cao and Liu [65] followed the re-assignment policies of [57, 62] for $2$-related machine setup. They considered re-assignment of last job of each machine and obtained overall competitive ratio of $min \{\sqrt{s+1}, \frac{s+1}{s}\}$ for different speed ratio($s$) intervals. 
We now present the main results obtained for semi-online scheduling on uniform related machines for the years 2006-2010 in table \ref{tab: Important Contributions for Uniform Machines: 2006-2010}. 
\begin{table}[h]
\caption {Main Results for Related Machines: 2006-2010}
\begin{tabular}{|c|p{3.8cm}|p{4.1cm}|}
\hline
\textbf{Author(s), Year} & \textbf{setup($\alpha| \beta| \gamma$)} & \textbf{Competitiveness Results} \\
\hline
Epstein and Ye 2007 [51] & $Q_2 |  LL | C_{min}$,       $Q_2 |  LL | C_{max}$ & $1.5$ UB and $1.465$ LB for $C_{max}$, $2.618$ Tight for $C_{min}$.  \\
\hline
Angelelli et al. 2008 [55] & $Q_2 | Sum | C_{max}$ & $1.33$ Tight for $x = 1$,  $x$ Tight for $x \in (1.28, 1.366)$, $\frac{x+2}{x+1}$ Tight for $x \geq 1.732$ \\
\hline
Englert et al. 2008 [56] & $P_m | re B(k) | C_{max}$,  $Q_m | re B(k) | C_{max}$  & ($1.333, 1.465$) LB and UB respectively for $P_m$ with $k \in \theta(m)$, ($2- \frac{1}{m-k+1}$) Tight for $P_m$ with $k \in [1, \frac{m+1}{2}]$, $2$ Tight for $Q_m$ with $k \in m$ \\
\hline
Chassid and Epstein 2008 [59] & $Q_2 | pmtn, GOS, Sum | C_{min}$ $Q_2 | pmtn, GOS, Sum | C_{max}$ & $1$ Tight for both setups \\ 
\hline
Ng et al. 2009 [60] & $Q_2 |  Opt | C_{max}$,  $Q_2 | Sum | C_{max}$  &  ($1.366, 1.369$) Tight with \textit{Opt} or \textit{Sum} respectively.\\
\hline
Ebenlendr and Sgall 2009 [61] &  $Q_3 | pmtn, Sum | C_{max}$    $Q_3 | pmtn, Max | C_{max}$  $Q_3 | pmtn, Decr | C_{max}$ & $1.138$ Tight with $S_1 = 1.414$, $S_2 = S_3 = 1$ and known \textit{Sum},  $1.252$ Tight with $S_1 =2$, $S_2 = S_3 = 1.732$ and known \textit{Max }, $1.52$ Tight with known \textit{Decr}. \\
\hline
Liu et al. 2009 [62] & $Q_2 | GOS | C_{max}$ $Q_2 | reasgn(last(k)) | C_{max}$ $Q_2 | reasgn(last)^{*} | C_{max}$ & ($1+ \frac{x+1}{x(2x+1)}$) LB with \textit{GOS} for $x>1$, ($1+\frac{1}{1+x}$) LB with reasgn(last(k)), ($\frac{(x+1)^2}{x+2}$) Tight with both re-assignment policies for $1\leq x\leq 1.414$ . \\ 
\hline
Angelelli et al. 2010 [63] & $Q_2 | Sum | C_{max}$ & $1.359$ LB with $b = 1.5$. \\
\hline
Cao and Liu 2010 [65] & $Q_2 | reasgn(last) | C_{max}$ & ($\sqrt{s}+1$) Tight for $1 \leq s < 1.618$, $\frac{s+1}{s}$ Tight for $s \geq 1.618$ \\
\hline
\end{tabular}
\label{tab: Important Contributions for Uniform Machines: 2006-2010}
\end{table}
\subsection{Recent Works in Semi-online Scheduling} \label{subsec:Recent Works in Semi-online Scheduling}
The recent era of semi-online scheduling has been dominated by non-preemptive scheduling in identical machines with multiple grades of service levels(\textit{GOS}) or machine hierarchy. Semi-online scheduling in \textit{unbounded batch machine} has been introduced. Several instances of related machines have been studied for various unexplored speed ratio intervals. Job rejection and reassignment policies have been introduced for various setups of related machines. We now present an overview of the state of the art in semi-online scheduling for unbounded batch machine, uniform related machines and identical machines as follows. \paragraph{}
\textbf{Unbounded Batch Machine:} Yuan et al. [68] introduced semi-online scheduling in single unbounded batch machine to improve the $1.618$ competitive bound obtained by pure online strategies in [25,32]. They considered that at any time step $t$, we are given with $p_t$ and $r_t$ of job $J_t$, where $J_t$ is the largest job that will arrive after time $t$. They obtained \textit{tight} bound $1.382$ with known $p_t$ by considering at most two batches. With given $r_t$, they achieved \textit{LB} $1.442$ and \textit{UB} $1.5$ by constructing at most three batches. With known $r_t$, they proposed an algorithm, which constructs at most two batches. The algorithm resets the value of $r_{t_1}$=$\max\{r_{t_1}, \alpha\cdot(p_{t_1})\}$, then forms the first batch ($U(t_1)$) by considering all jobs that are available by time $r_{t_1}$ and schedules them irrevocably on the machine, where $r_{t_1}$ is the release time of the first largest job and $\alpha$=$0.618$. The second batch $U(t_2)$ is formed by considering all jobs that are received at the time step $t_2$=$r_{t_1}+p_{t_1}$, then the value of $r_{t_2}$ is reset to $\max\{r_{t_2}, \alpha\cdot(p_{t_2})\}$ prior to schedule all jobs of batch $U(t_2)$. They obtained a matching \textit{UB} $1.618$ to that of pure online strategies. It is now a non-trivial challenge to beat the $1.618$ competitive bound by forming at most $2$ batches with known $r_t$.  \paragraph{}
\textbf{Related Machines:} \\
\textbf{Buffer.} Epstein et al. [95] investigated the setup $Q_m|re B(k)|C_{min}$ and proposed a $m$-competitive algorithm, where $m\geq 2$ and $k$=$m+1$. The algorithm keeps initial $m+1$ incoming jobs in the buffer. After arrival of the $(m+2)^{th}$ job until availability of the ${n}^{th}$ job, each time the smallest job $J_i$ is selected from $m+2$ available jobs and is scheduled by algorithm \textit{LS}, while not considering the machine speeds. When there is no jobs to arrive and the buffer contains $m+1$ jobs such that $p_1\leq p_2\leq ...\leq p_m\leq p_{m+1}$, the algorithm schedules  the jobs in any of the following rules.
\begin{enumerate}
\item Schedule $J_1$ by \textit{LS} rule and schedule the jobs $J_i$, where $2\leq i\leq (m+1)$ to the corresponding machine $M_j$ respectively, where $1\leq j\leq m$.
\item Schedule the jobs by rule $1$, but migrate $J_i$ to the machine $M_m$ for some $2\leq i\leq m$.
\item Schedule $J_{i+1}$ to $M_{i}$ for $1\leq i\leq m$ and schedule $J_1$ to a machine $M_k$ such that $2\leq k\leq (m-1)$.   
\end{enumerate}
Interestingly, the algorithm ignores the machine speeds until the arrival of all jobs, and then the relative order of the machines speeds are considered for making scheduling decision. Further, they studied the setup $Q_2|B(1)|C_{min}$, where $S_1$=$1$ and $S_2\geq 1$ and proposed a $\frac{2s+1}{s+1}$ competitive algorithm. The algorithm keeps the first job $J_1$ in the buffer, thereafter on the arrival of each incoming $J_i$, $2\leq i\leq n$, it is assumed that $p_x$=$\min\{p_{i-1}, p_i\}$ and $p_y$=$\max\{p_{i-1}, p_i\}$ . Now, job $J_x$ is scheduled on machine $M_1$ if $\frac{l_2+p_y}{s}\geq \frac{l_1+p_x}{s+1}$; otherwise $J_x$ is assigned to machine $M_2$. The goal is to schedule the smaller jobs to the slowest machine and relatively larger jobs to the fastest machine so as to maximize the minimum work load incurred on a machine. Lan et al. [76] studied the setup $Q_m|B(k)|C_{max}$ and achieved a tight bound ($2-\frac{1}{m}+\epsilon$) with $k$=$m$ and $m\geq 2$, where $\epsilon > 0$.\\
\textbf{Job Rejection.} Min et al. [96] initiated the study on semi-online scheduling in $2$-uniform machine with job rejection policy by considering $S_1$=$1$, $S_2\geq 1$. The rejection policy describes a scenario, where an incoming job $J_i$ can either be assigned to a machine or can be rejected permanently by incorporating a penalty of $x_i$. The objective of any semi-online algorithm is to incur a minimum value for the sum of makespan with sum of all penalties. The algorithm is given beforehand with two parallel processors for making scheduling policies, finally the best policy is opted for actual assignment of the jobs. Min et al. proposed a semi-online algorithm with the following rules for scheduling of each incoming job: Upon availability of a new job $J_i$, processor 1 rejects $J_i$, if $x_i\leq \alpha\cdot p_i$, where $\alpha$=$\frac{1}{s+1}$; else schedules $J_i$ by algorithm \textit{LS}. On the other hand, processor 2 rejects $J_i$, if $x_i\leq \beta\cdot p_i$, where $\beta$=$\frac{2}{2s+1}$; else schedules $J_i$ by algorithm \textit{LS}. After the assignments of all jobs, one of the policies that has yielded a minimum objective value is opted by the algorithm for actual scheduling of the jobs. The algorithm achieves tight bounds $\frac{2s+1}{s+1}$ for $1\leq s\leq 1.618$; and $\frac{s+1}{s}$ for $s> 1.618$. \\
\textbf{Max.} Cai and Yang [97] investigated the setup $Q_2|Max|C_{max}$ by considering $S_1$=$1$ and $S_2\geq 1$. They proposed algorithm \textit{Low Speed Machine Priority(LSMP)} and obtained tight bound $\max\{\frac{2s+2}{2s+1}, s\}$ for $s\in [1, 1.414]$. Algorithm \textit{LSMP} schedules an incoming job $J_i$ to machine $M_1$ if $p_i$=$p_{max}$; thereafter the remaining jobs are scheduled by algorithm \textit{LS}. If $p_i< p_{max}$, and if $l^{i}_{1}+p_{max}+p_i < l^{i}_{2}+\frac{p_i}{s}$, then schedules $J_i$ on $M_1$ (where, $l^{i}_{j}$ is the load of machine $M_j$ just before the scheduling of $J_i$); otherwise $J_i$ is scheduled on $M_2$. They proposed algorithm \textit{HSMP} and obtained tight bound $\frac{2s+2}{s+2}$ for $1\leq s\leq 1.414$. The tight bounds achieved for $s\geq 1.414$ are expressed by an algebraic function $r(s)$ as follows.
 \[
    r(s)= 
\begin{cases}
    \frac{s+2}{s+1},& \text{for}\hspace*{0.2cm} 1.414\leq s\leq 2\\
    \frac{3s+2}{2s+2},& \text{for}\hspace*{0.2cm} 2\leq s\leq 2.732\\
    \frac{s+1}{s},& \text{for}\hspace*{0.2cm} s\geq 2.732\\
\end{cases}
\]
The idea is to schedule the first largest job on machine $M_2$ and to schedule the remainning jobs by algorithm \textit{LS}.\\
\textbf{Opt.} Dosa et al. [69] followed the work of [33, 55, 60] for $Q_2|Opt|C_{max}$ by considering $S_1$=$1$, $S_2\geq 1$ and $s\in[1, 1.28]$. They obtained \textit{LB} $\min \{1+\frac{1}{3s}, 1+\frac{3s}{5s+5},  1+\frac{1}{2s+1}\}$. The \textit{LB} was derived by constructing a \textit{lower bound binary tree}, where each node represents a job $J_i$ along with its \textit{size} $p_i$ and each arc specifies an assignment of $J_i$ on machine $M_j\in\{M_1, M_2\}$. The left branch of a node represents scheduling of $J_i$ on $M_1$ and right branch specifies scheduling of $J_i$ on $M_2$. The \textit{size} of the next job $J_{i+1}$ is chosen based on its assignment to any of the $M_j$ in correspondence to the \textit{size} and scheduling of $J_i$. By traversing the \textit{lower bound binary tree} from root to the leaf nodes, one can obtain the instances, for which any semi-online algorithm achieves a \textit{CR} of at least the defined \textit{LB}. In [89, 94], the authors considered the setup studied in [69] and obtained lower bounds in terms of an algebraic function $r(s)$ for the following unexplored speed ratio intervals. 
\[
    r(s)= 
\begin{cases}
    \frac{6s+6}{4s+5},& \text{if}\hspace*{0.2cm} 1.3956\leq s\leq 1.443\\
    \frac{12s+10}{9s+7},& \text{if}\hspace*{0.2cm} 1.66\leq s\leq 1.6934\\
    \frac{18s+16}{16s+7},& \text{if}\hspace*{0.2cm} 1.6934\leq s\leq 1.6955\\
 \frac{8s+7}{3s+10},& \text{if}\hspace*{0.2cm} 1.6955\leq s\leq 1.6963\\ 
   \frac{12s+10}{9s+7},& \text{if}\hspace*{0.2cm} 1.6963\leq s\leq 1.7258\\ 
\end{cases}
\]
In [91], they studied for the interval $s \in[1.710, 1.732]$ and achieved tight bounds of $\frac{2s+10}{9s+7}$ for $s$=$1.7258$ and $\frac{s+1}{2}$ for $1.725\leq s\leq 1.732$ respectively. The obtained results draw an insight that a single algebraic function can not formulate the tightness of the \textit{LB}.\\
\textbf{Sum.} Dosa et al. [69] investigated the setup $Q_2|Sum|C_{max}$ by considering $S_1$=$1$, $S_2\geq 1$ and \textit{Sum}=$3s\cdot(1+s)$. They achieved \textit{tight} bounds  for the unexplored speed ratio interval $1\leq s< 1.2808$, which are presented by an algebraic function $r(s)$ as follows.
\[
    r(s)= 
\begin{cases}
    1+ \frac{1}{3s},& \text{for}\hspace*{0.2cm} s\in [1, 1.071]\\
   1+ \frac{3s}{4s+6},& \text{for}\hspace*{0.2cm} s\in [1.071, 1.0868]\\
    1+\frac{1}{2s+1},& \text{for}\hspace*{0.2cm} s\in [1.0868, 1.2808]\\ 
\end{cases}
\]
They proposed an algorithm by considering various time interval ranges as safe sets for scheduling decision making. The algorithm involves three subroutines as described below.\\
\textit{Subroutine 1 Master}
\begin{enumerate}
\item Upon the arrival of a new job $J_i$, run subroutine Slave.
\item If $J_i$=$J_1$, then run subroutine \textit{CoalA} from starting; else continue \textit{CoalA} from the breaking point of the last call for scheduling $J_{i-1}$.
\item If no more jobs to arrive, then stop; else move to step 1.
\end{enumerate}
\textit{Subroutine 2 Slave}
\begin{enumerate}
\item Schedule $J_i$ on machine $M_j$, if the value of $l_j+p_i$ is within the time interval range $[2s, 3s+1]$ for $j$=$1$ or $[3s^2-1, 3s^2+s] $ for $j$=$2$; thereafter, remaining jobs are scheduled on machine $M_{3-j}$ and stop.
\item Schedule $J_i$ on $M_j$, if $l_j\leq T_j$ and $l_j+p_j> T_{0j}$, where $T_1$=$3s+1-3s^2$, $T_2$=$s$, $T_{01}$=$3s+1$ and $T_{02}$=$3s^2+s$; thereafter, schedule the remaining jobs to machine $M_{3-j}$ and stop.
\item Schedule $J_i$ on $M_j$ if the value of $l_j+p_i$ is within the time interval range $[s-1, 3s+1-3s^2]$ for $j$=$1$ or the value within the range $[3s^2-s-2, s]$ for $j$=$2$; thereafter, schedule the remaining jobs on $M_{3-j}$ until $l_{3-j}< 2s$ for $j$=$2$ or $l_{3-j}< 3s^2-1$ for $j$=$1$; otherwise run the subroutine \textit{Slave} once more. 
\end{enumerate}
\textit{Subroutine 3 CoalA}
\begin{enumerate}
\item Schedule $J_i$ on $M_j$ until $l_j+p_i< s-1$.
\item Schedule $J_i$ on machine $M_1$, if $p_i< 3s^2-s-2$ and $l_1< 3s^2-2$.
\item Schedule $J_i$ on machine $M_2$, if $l_2+p_i< 3s^2-s-2$.
\item If $J_i$ is assigned to $M_2$, then the remaining jobs are scheduled on $M_2$ as long as $l_2\leq 3s^2-1$; thereafter, schedule the next job on $M_1$ and the remaining jobs on $M_2$; Stop.
\end{enumerate}
The algorithm was shown to be \textit{tight} with \textit{CR} of $1+\frac{1}{3s}$ for $s\in [1, 1.071]$. A similar algorithm with slight modification in the time interval ranges of the safe sets was proposed for $s\in [1.071, 1.08]$. \\
\textbf{GOS.} Hou and Kang [98] invesigated semi-online hierarchical scheduling on $m$-uniform machine setup with \textit{max-min} and \textit{min-max} objectives. They considered $m$ machines in two hierarchies, where in hierarchy $1$, we have $k$ machines, each with speed $S_1\geq 1$ and $g_j$=$1$, capable of executing all jobs. In hierarchy $2$, we have rest $m-k$ machines, each with speed $S_2$=$1$ and $g_j$=$2$, capable of executing the jobs having $g_i$=$2$. For $Q_m|GOS|C_{min}$, they proved that no online algorithm can be possible with a bounded \textit{CR}. They investigated the setup $Q_m|pmtn, GOS|C_{min}$ and obtained \textit{UB} of $\frac{2ks+m-k}{ks+m-k}$ for $0< s< \infty$ by applying the fractional assignment policy, where each incoming job $J_i$ can be splitted arbitrarily among the machines. For $Q_m|pmtn, GOS|C_{max}$, they achieved \textit{UB} of $\frac{(ks+m-k)^2}{k^2s^2+ks(m-k)+(m-k)^2}$ for $0< s< \infty$. For $Q_m|pmtn, GOS, Sum|C_{max}$, they proposed a $1$-competitive algorithm. The idea is to schedule the jobs having $g_i$=$1$ evenly on the machines having $g_j$=$1$ and schedule the jobs having $g_i$=$2$ on the machines with $g_j$=$2$ as long as the loads of the machines are under a given threshold value.   
Lu and Liu [81] studied three variants of $Q_2|GOS|C_{max}$ with known Opt or Sum or Max by considering $S_1$=$1$, $S_2$=$s>0$, $g_i$=$1$(for processing job $J_i$ exclusively on machine $M_1$) and $g_i$=$2$(for making $J_i$ eligible for processing in any one of the $M_j\in\{M_1, M_2\}$). They proposed algorithm \textit{Gos-OPT} for $Q_2|GOS, Opt|C_{max}$ and obtained \textit{UB} of $\min\{\frac{1+2s}{1+s}, \frac{1+s}{s}\}$. Algorithm \textit{Gos-OPT} schedules each incoming job $J_i$ on the machine $M_1$ if $g_i$=$1$; else if $g_i$=$2$ then $J_i$ is scheduled by the following policy: if $s\geq \frac{1+\sqrt{5}}{2}$, then $J_i$ is scheduled on machine $M_2$; if $0<s<\frac{1+\sqrt{5}}{2}$ and $\frac{l_2+p_i}{s}\leq (\frac{1+2s}{1+s})\cdot C_{OPT}$, then $J_i$ is scheduled on $M_2$; if $0<s<\frac{1+\sqrt{5}}{2}$ and $\frac{l_2+p_i}{s}> (\frac{1+2s}{1+s})\cdot C_{OPT}$, then $J_i$ is scheduled on $M_1$. They achieved a matching \textit{UB} for $Q_2|GOS, Sum|C_{max}$ with an equivalent algorithm \textit{Gos-SUM}, which is equivalent to algorithm \textit{Gos-OPT}, just simply replacing $C_{OPT}$ by  $\frac{Sum}{1+s}$ in the policy. They proposed algorithm \textit{GoS-MAX} for $Q_2|GOS, Max|C_{max}$ and obtained \textit{UB} $1+s$ for $0<s<\frac{\sqrt{5}-1}{2}$. Algorithm \textit{GoS-MAX} schedules an incoming job $J_i$ on $M_1$ if $g_i$=$1$; else if $g_i$=$2$, then $J_i$ is scheduled by the following policy: if $0<s\leq \frac{\sqrt{5}-1}{2}$, then $J_i$ is scheduled on $M_1$; if $\frac{\sqrt{5}-1}{2}<s\leq 1$ and $\frac{l_2+p_i}{s}\leq \frac{1+\sqrt{5}}{2}\cdot (\max\{p_{max}, \frac{T_k}{1+s}, T^{k}_{1}\}$), then $J_i$ is scheduled on $M_1$; if $1<s<s^*$ and $\frac{l_2+p_j}{s}\leq (\frac{1+\sqrt{1+4s}}{2}\cdot (\max\{\frac{p_{max}}{s}, \frac{T_k}{s}, T^{k}_1\}))$, then $J_i$ is scheduled on $M_2$, else $J_i$ is assigned to machine $M_1$; and if $s\geq s^*$, then $J_i$ is scheduled on $M_2$. They achieved \textit{UB} $\min\{1+s, \frac{1+\sqrt{5}}{2}\}$ for $0<s<1$ and \textit{UB} $\min\{\frac{1+\sqrt{1+4s}}{2}, \frac{1+s}{s}\}$ for $s\geq 1$. (Note that: $T_k$ is referred to as sum of the sizes of the first $k$ jobs, $T^{1}_{k}$ represents the sum of the sizes of the set of jobs belongs to first $k$ jobs for which $g_i$=$1$ and $s^*\in (1.3247, 1.3248)$). \\
\textbf{TGRP.}  Luo and Xu [88] followed the work of Chassid and Epstein [59] for semi-online scheduling on $2$-parallel machines with \textit{Max-Min} objective. They considered hierarchical scheduling to cater different levels of services to the input jobs. They investigated two semi-online variants, one with \textit{TGRP}[$1, b$] and obtained lower bound of $1+b$ for $b \geq 1$. In the second case, they considered   \textit{TGRP}[$1, b$], \textit{Sum} and achieved lower bound of $b$ for $1 \leq b < 2$. Cao and Liu [99] studied the setup $Q_2|TGRP|C_{max}$ by considering $S_1$=$1$, $S_1\leq S_2$=$s$ and $\forall_{i}$, $p_i\in [p, xp]$, where $p> 0$ and $x\geq 1$. They proved tight bound of $\min\{\frac{2s+1}{s+1}, \frac{s+1}{s}, x\}$ for algorithm \textit{LS}, where $1\leq s\leq 1.618$ and $x\geq \frac{1+s}{s{\alpha}^2}$ with $\alpha$=$\frac{1+s-s^2}{s^2}$. They proposed a $s$-competitive algorithm, which is tight for $1.325\leq s\leq 1.618$ and $x\leq \frac{s^2-1}{1+s-s^2}$. The algorithm schedules an incoming job $J_i$ on machine $M_2$ if $l^{i}_{2}+\frac{p_i}{s}\leq s\cdot \frac{l^{i}_{1}+s\cdot l^{i}_{2}+p_i}{1+s}$; otherwise schedules $J_i$ on machine $M_1$. Further, they designed a new algorithm that achieves the following tight bounds for the unexplored speed ratio intervals, expressed in an algebraic function $r(s)$.
 \[
    r(s)= 
\begin{cases}
    s,& \text{for}\hspace*{0.2cm} s\in [1.206, 1.5]\hspace*{0.2cm} and \hspace*{0.2cm} s\leq x\leq \min\{2s-1, \frac{2s^2-2}{1+s-s^2}\}\\
   \frac{1+x}{2},& \text{for}\hspace*{0.2cm} s\in [1, 1.28]\hspace*{0.2cm} and \hspace*{0.2cm}\max\{2s-1, \frac{-s+\sqrt{9s^2+8s}}{2s}\}\leq x\leq \frac{2}{s}\\
\end{cases}
\]
The new algorithm schedules the initial four incoming jobs $J_1$, $J_2$, $J_3$ and $J_4$ on machines $M_2$, $M_1$, $M_2$ and $M_1$ respectively. Thereafter, each incoming $J_i$, where $i\geq 5$ is scheduled on machine $M_2$ if $l^{i}_{2}+\frac{p_i}{s}\leq \frac{k\cdot (l^{i}_{1}+sl^{i}_{2}+p_i)}{1+s}$, where $k$=$\max\{s, \frac{1+x}{2}\}$; else $J_i$ is scheduled on machine $M_1$. (Note that: $l^{i}_{j}$ is the load of machine $M_j$ just before the scheduling of job $J_i$.) \\
\textbf{Job Reassignment.}  Englert et al. [93] initiated the study on online scheduling in $m$-uniform machine with job reassignment policy, where $m\geq 2$. The aim is to explore, how far the reassignment of a small number of jobs helps to improve the \textit{CR} of an algorithm designed for makespan minimization in pure online setup. They proposed an algorithm that achieves a \textit{CR} between $1.33$ and $1.7992$ for different speed ratio intervals with at most $m$ reassignments. The algorithm functions in two phases, where in the first phase, online arriving jobs are scheduled on $m$ machines. And in the second phase, a specified number of jobs(at most $m$) are removed from the allocated machines and are re-scheduled on different machines as per a defined set of rules. We now present a summary of the main results obtained in recent years for semi-online scheduling on related machines in table \ref{tab: Important Results for Uniform Machines: 2011-2017}.         
\begin{table}[h]
\caption {Summary of the Recent Contributions for Related Machines}
\begin{tabular}{|c|p{3.9cm}|p{4.9cm}|}
\hline
\textbf{Author(s), Year} & \textbf{setup($\alpha| \beta| \gamma$)} & \textbf{Competitiveness Results} \\
\hline
Epstein et al. 2011 [95] & $Q_2 | re B(k) | C_{min}$ $Q_m | re B(k)| C_{min}$ & $\frac{2s+1}{s+1}$ Tight for $Q_2$ and $m$ Tight for $Q_m$ \\
\hline
Cai and Yang 2011 [97] & $Q_2 | Max | C_{max}$ & $\frac{2s+2}{2s+1}$ for $s\in [1, 1.414]$; $\frac{s+1}{s}$ for $s\in [1.414, 2.732]$\\
\hline
Dosa et al. 2011 [69] & $Q_2 | Opt | C_{max}$ $Q_2 | Sum | C_{max}$ & min$\{1+\frac{1}{3s}, 1+\frac{3s}{5s+5}, 1+\frac{1}{2s+1}\}$ LB with \textit{Opt}, ($1+\frac{1}{3s}$) Tight with  \textit{Sum}. \\ 
\hline
Hou and Kang 2011 [98]  & $Q_m|pmtn, GOS|C_{min}$  $Q_m|pmtn, GOS, Sum|C_{max}$ & $\frac{2ks+m-k}{ks+m-k}$ UB for $C_{min}$, $1$ Tight for $C_{max}$\\
\hline 
Lan et al. 2012 [76] & $Q_m|B(k)|C_{max}$ & ($2-\frac{1}{m}+\epsilon$) Tight with $k$=$m$\\
\hline  
Lu and Liu 2013 [81] & $Q_2 |  GOS, Opt | C_{max}$ $Q_2 |  GOS, Sum | C_{max}$ $Q_2 |  GOS, Max | C_{max}$ & $min \{\frac{1+2s}{1+s}, \frac{1+s}{s}\}$ Tight with \textit{Opt} or \textit{Sum}, $min \{\frac{1+ \sqrt{1+4s}}{2}, \frac{1+s}{s}\}$ Tight with \textit{Max} for $s \geq 1$. \\ 
\hline
Luo and Xu 2015 [88] & $Q_2 | GOS, TGRP | C_{min}$ $Q_2 | GOS, TGRP, Sum | C_{min}$ & ($1+b$) LB with \textit{TGRP},   ($b$) LB with \textit{TGRP} and \textit{Sum}. \\ 
\hline
Dosa et al. 2015 [89] & $Q_2 | Opt | C_{max}$ & ($\frac{6(s+1)}{4s+5}$) LB for $s \in [1.3956, 1.443]$, $min \{\frac{12s+10}{9s+7}, \frac{18s+16}{16s+7}, \frac{8s+7}{3s+10}, \frac{12s+10}{9s+7}\}$ LB for $s \in [1.666, 1.725]$. \\
\hline
Cao and Liu 2016 [99] & $Q_2|TGRP|C_{max}$ & $s$ Tight for $s\in[1.325, 1.618]$\\
\hline
Dosa et al. 2017 [91] & $Q_2 | Opt | C_{max}$ & ($\frac{12s+10}{9s+7}$) Tight for $s \approx 1.7258$, ($\frac{s+1}{2}$) Tight for $1.725 \leq s < 1.732$. \\
\hline
Englert et al. 2018 [93] & $Q_m | reasgn | C_{max}$ & $1.7992$ UB \\
\hline
\end{tabular}
\label{tab: Important Results for Uniform Machines: 2011-2017}
\end{table}
\paragraph{}
\textbf{Identical Machines:}\\
\textbf{GOS.} Liu et al. [73] studied the setup $P_2|GOS, TGRP[a, ba]C_{max}$, where $a> 0$ and $b> 1$. The \textit{GOS} model studied here considers two machines, where one machine can afford higher quality in service called \textit{higher GOS machine} and the other one can cater normal quality in service called \textit{lower GOS machine}. Each newly available job $J_i$ reveals its $p_i$ and $g_i$, where $g_i\in \{1, 2\}$. If $g_i$=$1$, then $J_i$ must be executed on machine $M_1$; if $g_i$=$2$, then $J_i$ can be executed on any of the machines. They proposed the algorithm \textit{B-ONLINE} by following the the policy of Park et al. [46] and obtained the following \textit{tight} bounds, expressed by an algebraic function $r(s)$.
 \[
    r(s)= 
\begin{cases}
    \frac{1+b}{2},& \text{for}\hspace*{0.2cm} 1.785\leq b\leq 2\\
   1.5,& \text{for}\hspace*{0.2cm} 2\leq b\leq 5\\
   \frac{4+b}{6},& \text{for}\hspace*{0.2cm} 5\leq b\leq 6\\
\end{cases}
\]
Algorithm \textit{B-ONLINE} works by the following policy. Initialize the parameters $l_1$=$0$, $l_2$=$0$, $P_{max}$=$0$, $X$=$0$ and $T$=$0$. Upon receiving a new job $J_i$, update $P_{max}$=$\max\{P_{max}, p_i\}$ and $X$=$X+\frac{p_i}{2}$. If $g_i$=$1$, then schedule $J_i$ on machine $M_1$ and update $T$=$T+p_i$. If $g_i$=$2$, then $J_i$ is scheduled on $M_2$ if $l_2+p_i\leq r(s)\cdot L$, where $L$=$\max\{X, T, P_{max}\}$; otherwise, $J_i$ is assigned to machine $M_1$. Further, they studied the setup $P_2|GOS, TGRP[a, ba], Sum|C_{max}$ and proposed the ($\frac{1+b}{2}$)-competitive optimal algorithm \textit{B-SUM-ONLINE} for $Sum\geq (\frac{2b}{b-1})\cdot a$ and $1< b< 2$. Algorithm \textit{B-SUM-ONLINE} schedules an incoming job $J_i$ on $M_1$ if $g_i$=$1$. If $g_i$=$2$ and $l_2+p_i\leq (\frac{1+b}{2})\cdot L$, then $J_i$ is scheduled on $M_2$; otherwise $J_i$ is scheduled on $M_1$.  
Wu et al. [77] followed the work of Liu et al. in the study of the setup $P_2|GOS, Opt|C_{max}$ and proposed a $1.5$ competitive optimal algorithm. The algorithm schedules an incoming job $J_i$ on machine $M_1$ if $g_i$=$1$. If $g_i$=$2$ and $l_2+p_i\leq (1.5)\cdot C_{OPT}$, then $J_i$ is scheduled on $M_2$; otherwise $J_i$ is assigned to machine $M_1$. The objective is to keep the loads of both machines under $(1.5)\cdot C_{OPT}$. Further, they considered \textit{GOS} with known $p_{max}$ in $2$-identical machine setup and proposed $1.618$ competitive optimal algorithm \textit{Gos-Max}. Algorithm \textit{Gos-Max} works as follows: upon receiving a job $J_i$, update $X$=$X+\frac{p_i}{2}$, where $X$=$0$. If $g_i$=$1$, schedule $J_i$ on $M_1$ and update $T$=$T+p_i$, where, $T$=$0$. If $g_i$=$2$ and $l_2+p_i\leq (1.618)\cdot L$, then $J_i$ is scheduled on $M_2$, where, $L$=$\max\{p_{max}, X, T\}$; else schedule $J_i$ on machine $M_1$.\\
Chen et al. [80] studied the setup $P_2|GOS, B(k)|C_{max}$ by considering known $g_i\in \{1, 2\}$ for each incoming job $J_i$. It is assumed that machine $M_1$ can execute all jobs, whereas machine $M_2$ can execute the jobs having $g_i$=$2$. They proposed a $1.5$ competitive optimal algorithm, which always tries to keep the largest job having $g_i$=$2$ in the buffer and schedule it at the end. The algorithm works in two phases, wherein the first phase, all jobs having $g_i$=$1$ are scheduled on $M_1$ and maximum possible jobs are assigned to $M_2$ as long as the desired \textit{CR} holds. In the $2^{nd}$ phase, the largest job in the buffer is scheduled on the smallest loaded machine. Further, they studied the setup $P_2|GOS, reasgn(k)|C_{max}$ and proposed a $1.5$ competitive optimal algorithm with $k$=$1$. The idea is to schedule maximum number of jobs on a particular machine $M_j$ until $l_j$ reaches upto a defined threshold, then reassign the largest job scheduled on $M_j$ to the other machine $M_{3-j}$. Zhang et al. [82] improved the bounds obtained by Liu et al. in [73] with \textit{GOS} and \textit{TGRP}($a, ba$) for $1 \leq b < 3$. Further, they proved that use of preemption and idle time do not improve the competitiveness of the pure online setting of hierarchical scheduling in $2$-identical machines.
Luo and Xu [85] improved the bounds given in [46, 77] for $2$-identical machines with known \textit{Sum} and different \textit{GOS} levels such as higher \textit{GOS} and lower \textit{GOS}. \\
Chen et al. [100] extended their previous work [80] with similar idea and considered three different setups of online hierarchical scheduling in $2$-identical machines. They studied the setup, where $\sum_{}^{}{p_i}$ for the jobs with $g_i$=$1$ is known and proved a \textit{tight} bound $1.5$ for algorithm \textit{LS}. In another setup, they assumed known values of $T_1$=$\sum_{}^{}{p_i}$=$1$, $\forall J_i$ such that $g_i$=$1$ and the value of $T_2$=$\sum_{}^{}{p_i}$=$T> 0$, $\forall J_i$ such that $g_i$=$2$. They proposed the algorithm \textit{CMF}, which achieves a \textit{tight} bound of $1.33$. Algorithm \textit{CMF} adopts the following rule: schedule an incoming job $J_i$ by its $g_i\in \{1, 2\}$ to respective $M_j\in \{M_1, M_2\}$ if $T\leq 2$. If $T> 2$, else if $g_i$=$1$, then $J_i$ is scheduled on machine $M_1$, else if $g_i$=$2$, then $J_i$ is assigned to $M_1$ if $l^{i}_{1}+1+p_j\leq \frac{1+T}{3}$, else let $x$=$i$, schedule $J_x$ and the remaining jobs by following rule: assign $J_x$ to $M_2$ and remaining jobs to $M_1$ if $l^{x}_{1}+1+p_x> \frac{2(1+T)}{3}$, else $J_x$ along with all future jobs for which $g_i$=$1$ are scheduled on $M_1$ and rest of the jobs are assigned to machine $M_2$. Further, they considered $B(k)$ in the first setup and obtained a \textit{tight} bound of $1.33$ with $k$=$1$.\\
Qi and Yuan [101] addressed the research challenge posed by Chen et al. in [80] regarding an unified approach for semi-online hierarchical scheduling with buffer or reassignment. However, they opened up a new direction by introducing \textit{$L_p$-norm load balancing($C^{(p)}$)} as an optimality criterion for semi-online hierarchical scheduling in $2$-identical machine setup. Let us represent in a schedule the final loads of $M_1$ and $M_2$ by $l_1$ and $l_2$ respectively. The load vector is represented by $L$=$\{l_1, l_2\}$. The $L_p$-norm is denoted as $\|L_p\|$ and defined as follows:  \[
    \|L_p\|= 
\begin{cases}
    (l^{p}_{1}+l^{p}_{2})^{\frac{1}{p}},& \text{for}\hspace*{0.2cm} 1\leq p< \infty\\
   \max\{l_1, l_2\},& \text{for}\hspace*{0.2cm} p=\infty\\
\end{cases}
\]
They argued that $L_p$-norm objective is practically more significant than makespan, as it captures the average machine loads instead of the largest load among the machines. They obtained \textit{tight} bound $1.5$ by separately considering $B(k)$ and $reasgn(k)$ respectively with $k$=$1$ for $p$=$\infty$.
Xiao et al. [102] followed the work of Chen et al. [100] and addressed $C_{min}$ objective in a setting, where \textit{sum of the sizes of low hierarchy jobs($T_1$=$1$)} is known and $B(1)$ is given. They proposed the algorithm \textit{BLS}, which achieves a \textit{tight} bound $1.5$ for $C_{min}$. Algorithm \textit{BLS} schedules an incoming job $J_i$ on $M_1$ if $g_i$=$1$. If $g_i$=$2$, put $J_i$ on the buffer, $\Big($let $B_{max}$=$\max\{p_i|jobs\hspace*{0.1cm} in \hspace*{0.1cm} the\hspace*{0.1cm} buffer\}$,  $B_{min}$=$\min\{p_i|jobs\hspace*{0.1cm} in \hspace*{0.1cm} the\hspace*{0.1cm} buffer\}\Big)$ and if $l_2+B_{max}\geq \frac{l_1+T_1+B_{min}}{2}$, then $J_{min}$ is scheduled on $M_1$; else, $J_{min}$ is assigned to machine $M_2$. Further, they considered the setting, where $T_1$ is given and \textit{$p_{max}$} for hierarchy 2 i.e. $p^{2}_{max}$  is known. They obtained a \textit{tight} bound $1.5$ for $C_{min}$. Qi and Yuan [103] studied the setup $P_2|GOS, Sum|C^{(p)}$ and proposed an algorithm that achieves a \textit{tight} bound $1.5$ for $p$=$\infty$. The algorithm schedules an incoming $J_i$ on machine $M_i$ if $g_i$=$1$. If $g_i$=$2$, and $l_2+p_i\leq \frac{3}{4}\cdot T$, then $J_i$ is assigned to machine $M_2$. If $g_i$=$2$ and $l_2+p_j> \frac{3}{4}\cdot T$, then schedules $J_i$ by the following rule: if $l_2< \frac{1}{4}\cdot T$ and $l_2\leq \frac{T-p_i}{2}$, then schedule $J_i$ on $M_2$ and all future jobs on $M_1$, If $l_2< \frac{1}{4}\cdot T$ and $l_2> \frac{T-p_i}{2}$, then schedule $J_i$ on $M_1$ and all future jobs with $g_i$=$2$ on $M_2$ and jobs with $g_i$=1 on $M_1$. If $l_2\geq \frac{1}{4}\cdot T$, then schedule $J_i$ along with all future jobs on machine $M_1$. The idea is to schedule larger number of jobs on $M_2$ as long as $l_2$ would not exceed $l_1$. Importantly, the algorithm handles the larger size jobs with known $T$. Further, they studied the setup $P_2|GOS, T_1, T_2|C^{(p)}$ and obtained a \textit{tight} bound of $1.33$ for $p$=$\infty$. 
In future, some interesting consideration would be semi-online hierarchical scheduling for $L_p$-norm optimization with other unexplored \textit{EPIs} such as \textit{Max}, \textit{TGRP}, \textit{Opt}, \textit{Decr} etc. The study remains open in $m$-identical machine for $m>2$ and related machine setups. We now present the summary of important results for semi-online scheduling in identical machines with \textit{GOS} in table \ref{tab: Important Results for Identical Machines with GOS: 2011-2017}.
\begin{table}[h]
\caption {Summary of the Recent Works on Identical Machines with GOS}
\begin{tabular}{|c|p{4.1cm}|p{4.7cm}|}
\hline
\textbf{Author(s), Year} & \textbf{setup($\alpha| \beta| \gamma$)} & \textbf{Competitiveness Results} \\
\hline
Liu et al. 2011 [73] & $P_2 | GOS, TGRP | C_{max}$ $P_2 | GOS, TGRP, Sum | C_{max}$ & max$\{\frac{1+b}{2}, 1.5, \frac{4+b}{6}\}$ Tight with GOS and TGRP for $1 < b < 6$, $\frac{1+b}{2}$ Tight with GOS, TGRP and Sum for $1 < b < 2$. \\
\hline
Wu et al. 2012 [77] & $P_2 | GOS, Opt | C_{max}$ $P_2 | GOS, Max | C_{max}$ & ($1.5, 1.618$) Tight for respective setups. \\
\hline
Chen et al. 2013 [80] & $P_2 | GOS, B(k) | C_{max}$,  $P_2 | GOS, reasgn(k) | C_{max}$ & $1.5$ Tight with $k=1$ for both setups \\
\hline
Zhang et al. 2013 [82] & $P_2 |  GOS, TGRP | C_{max}$ $P_2 |  pmtn, GOS, TGRP | C_{max}$ & $1.66$ Tight with $b \geq 3$ for N-pmtn, $1.5$ Tight with $b \geq 2$ for pmtn. \\
\hline
Luo and Xu 2014 [85] &  $P_2 | GOS, Sum | C_{max}$ &  ($1.5, 1.53, 1.33$) Tight for \textit{Sum} with (higher GOS or lower GOS or both) respectively.\\
\hline
Chen et al. 2015 [100] & $P_2|GOS, Sum|C_{max}$ $P_2|GOS, Sum, B(1)|C_{max}$ & $1.33$ Tight for respective setups\\
\hline
Qi and Yuan 2016 [101] & $P_2|GOS, B(1)|C^{(p)}$ $P_2|GOS, reasgn(1)|C^{(p)}$ & $1.5$ Tight for both setups for $p$=$\infty$\\
\hline
Xiao et al. 2019 [102] & $P_2|GOS, T_1, B(1)|C_{min}$ $P_2|GOS, T_1, p^{2}_{max}|C_{min}$ &  $1.5$ Tight for both setups\\
\hline
Qi and Yuan 2019 [103] & $P_2|GOS, Sum|C^{(p)}$ & $1.5$ Tight for $p$=$\infty$\\
\hline
\end{tabular}
\label{tab: Important Results for Identical Machines with GOS: 2011-2017}
\end{table} \\
\textbf{TGRP.} Cao et al. [72] studied the setup $P_2|TGRP, Max|C_{max}$ by considering $TGRP[a, ba]$ and $p_{max}$=$ba$, where $a> 0$ and $b\geq 1$. They obtained $\frac{b+1}{2}$ \textit{LB} for $1\leq b< 1.33$ and $1.33$ \textit{LB} for $b\geq 2$. They proposed algorithm \textit{PIJS}, which achieves a \textit{tight} bound $\max\{\frac{4(b+1)}{3b+4}, \frac{2b}{b+1}\}$ for $1.33\leq b\leq 2$. Algorithm \textit{PIJS} schedules an incoming job $J_i$ on machine $M_1$ if $l^{i}_{1}+p_i\leq k\cdot \max\{q^{1}_{i}+q^{2}_{i}+...+q^{\lceil\frac{i+1}{2}\rceil}, \frac{l^{i}_{1}+l^{i}_{2}+p_i+p_{max}}{2}\}$; otherwise $J_i$ is scheduled on $M_2$. And this continues until the arrival of the first largest job (let $J_{max}$). When $J_{max}$ arrives, it is scheduled on machine $M_2$. Thereafter, each incoming $J_i$ is scheduled on $M_1$, if $l^{i}_{1}+p_i\leq k\cdot \max\{q^{1}_{i}+q^{2}_{i}+...+q^{\lceil\frac{i}{2}\rceil}, \frac{l^{i}_{1}+l^{i}_{2}+p_i}{2}\}$; else $J_i$ is scheduled on machine $M_2$. (Note that: $k$=$\max\{\frac{4(b+1)}{3b+4}, \frac{2b}{b+1}\}$, $l^{i}_{j}$ is the load of machine $M_j$ just before the assignment of $J_i$ and $q^{r}_{i}$ is the $r^{th}$ smallest job at the arrival of $J_i$ i.e. $\{q^{1}_{i}, q^{2}_{i},...,q^{r}_{i}\}$ such that $p^{1}_{i}\leq p^{2}_{i}\leq...\leq p^{r}_{i}$.) The idea given in this study reveals that when $p_{max}$ is known in advance, it is better to assign $J_{max}$ at the outset. Cao and Wan [84] studied the setup $P_2|TGRP[1, b], Decr|C_{max}$. They showed that algorithm \textit{LS} achieves a \textit{tight} bound $1.16$ for $1\leq b\leq 1.5$. With only known $TGRP(ub)$, they obtained \textit{LB} $1.16$, which matches the \textit{UB} given by Seiden in [21], where, $ub\geq 1.5$.\\
\textbf{Arrival Order of Jobs.} Li et al. [70] studied the scenario where an incoming job $J_i$ requests an order to the scheduler with its release time $r_i$ and processing time $p_i$. Then, the scheduler service the order by non-preemptively schedule the job with the objective to optimize the makespan. They considered that jobs are arriving by non-decreasing release times(\textit{Incr-r}) and non-increasing sizes(\textit{Decr}). They analyzed the performance of algorithm \textit{LS} and obtained \textit{UB} $\frac{3}{2}-\frac{1}{2m}$ for $m$-identical machine setup. Cheng et al. [78] studied the setup $P_m|Decr|C_{max}$. They analyzed algorithm \textit{LPT} and proved \textit{tight} bounds $1.18$ for $m$=3 and $1.25$ for $m>3$. Tang and Nai [87] refined the results of Li et al. [70] and derived a new proof for the \textit{UB} of algorithm \textit{LS}. \\
\textbf{Job Reassignment.}  Min et al. [71] considered the job's assignment policy of Tan and Yu [57]. They obtained a tight bound of $1.41$ for semi-online scheduling on $2$-identical machine by allowing the reassignment of the last job of one machine only. Further, they considered known \textit{Sum} besides the reassignment policy and improved the previous best \textit{UB} of $1.33$ to $1.25$. \\
\textbf{Combined Information.} Cao et al. [74] considered several semi-online variants for scheduling on $2$-identical machine with \textit{min-max} objective. They proposed $1.2$ competitive optimal algorithm \textit{OM} with known \textit{Opt} and \textit{Max}. Further, they considered combined information on    (\textit{B(k), Max}), (\textit{B(1), Decr}), (B(1), TGRP(1,b)) and obtained \textit{tight} bounds ($1.25, 1.16, 1.33$) respectively. Lee and Lim [79] studied the setup $P_m|Sum, Max| C_{max}$ and achieved \textit{UB}s ($1.462$, $1.5$) for $m$=$4$, $5$ respectively. \\
\textbf{Sum.} Albers and Hellwig [75] studied the setup $P_m|Sum|C_{max}$. They improved the \textit{LB} $1.565$ [35] to $1.585$ for $m$-identical machine setup, where $m\rightarrow \infty$. They proposed algorithm \textit{Light Load}, which is free from traditional job class policy considered in [35, 44] and achieves an \textit{UB} $1.75$. Lee and Lim [79] investigated the setup $P_m|Sum|C_{max}$ for small number of machines. They obtained \textit{LB}s of $1.442$, $1.482$ and $1.5$ for $m$=$4$, $5$ and $6$ respectively. An algorithm named \textit{ForwardFit-BackwardFit-ListScheduling} was proposed by assuming $Sum$=$m$, which achieves \textit{UB}s of $1.4$, $1.4615$ and $1.5$ for $m$=$3$, $4$ and $5$ respectively. The algorithm prefers two conditions, where in the first condition, a \textit{load threshold} is set upto the defined competitive bound to keep the loads of the machines under the threshold value.  Before scheduling an incoming job, the loads of each machine is checked against the load threshold value. If the first condition fails, then the incoming job is scheduled by algorithm \textit{LS}.     An obvious question raised here is: How to choose the threshold value, which always guarantees that the scheduling of all jobs would yield the defined competitive bound for any $m$?\\
Kellerer et al. [90] obtained an \textit{UB} $1.585$ by considering $Sum$=$m$, which matches the \textit{LB} achieved by Albers and Hellwig in [75] for $m$-identical machine setup. They adopted the job class policy and classified the incoming jobs into four classes such as tiny, small, medium and large depending on their sizes, defined by the time intervals $(0, \frac{\alpha}{2}]$ for tiny, $(0, \alpha]$ for small, $(\alpha, \frac{1}{2\alpha}]$ for medium, $[>\frac{1}{2\alpha}]$ for large. Similarly, $m$ machines were classified as tiny, small, medium, big and huge depending on their loads defined by the time intervals $(0, \frac{\alpha}{2}]$ for tiny, $(0, \alpha]$ for small, $(\alpha, \frac{1}{2\alpha}]$ for medium, $(\frac{1}{2\alpha}, 1]$ for big and $[> 1]$ for huge, where $\alpha$=$0.585$. They proposed an algorithm, which executes in 2 phases, where in the $1^{st}$ phase, jobs are scheduled on the machines depending on the classes of jobs and machines. The $2^{nd}$ phase of the algorithm, emerges from the $1^{st}$ phase and runs two policies with respect to the current machines loads after the $1^{st}$ phase. Here, the classification of jobs and machines helps in improving the tightness in the competitive bound to $1+\alpha$. A natural question pops out from here is: Can an algorithm be possible for $P_m|Sum|C_{max}$ with job class policy and $\alpha< 0.585$? \\
\textbf{Buffer.} Lan et al. [76] studied the general cases for identical machines by considering buffer as additional feature. For $m$-identical machines they obtained a tight bound $1.5$ with a buffer of size $1.5m$. 
\\
\textbf{Opt.} Kellerer and Kotov [104] studied the setup considered by Azar and Regev in [15]. They improved the \textit{UB} from $1.625$ to $1.571$ by considering the job class policy, where the incoming jobs and available machines are classified by their sizes and current loads respectively, defined by the specified time interval ranges. It was proved that a two phase algorithm with job class policy always guarantees the loads of each machine to be under $1.571$ of the known $Opt$. Gabay et al. [105] further improved the \textit{UB} to $1.5294$. The current best known \textit{UB} $1.5$ for the setup $P_m|Opt|C_{max}$ is due to Bohm et al. [106]. Further, they obtained UB $1.375$ for $m$=$3$. Gabay et al. [107] improved the  \textit{LB} $1.33$ to $1.357$ for the setup $P_m|Opt|C_{max}$. Thus, minimizing the gap between the current best \textit{LB} and \textit{UB} in $m$-identical machine semi-online scheduling with known \textit{Opt}. We now present the summary of important results obtained for semi-online scheduling in identical machines other than known \textit{GOS} in table \ref{tab: Important Results for Identical Machines: 2011-2017}.
\begin{table}[h]
\caption {Summary of the Recent Works on Identical Machines}
\begin{tabular}{|c|p{4.5cm}|p{4cm}|}
\hline
\textbf{Author(s), Year} & \textbf{setup($\alpha| \beta| \gamma$)} & \textbf{Competitiveness Results} \\
\hline
Cao et al. 2011 [72] & $P_2 | TGRP, Max | C_{max}$ & $\frac{b+1}{2}$ LB for $1 \leq b < 1.33$, $1.33$ LB for $b \geq 2$, $\max\{\frac{4(b+1)}{3b+4}, \frac{2b}{b+1}\}$ Tight for $1.33 \leq b < 2$.   \\
\hline
Li et al. 2011 [70] & $P_m | Incr-r, Decr | C_{max}$ & ($\frac{3}{2} - \frac{1}{2m}$) Tight. \\
\hline
Min et al. 2011 [71] &  $P_2 | reasgn(last(1)^*) | C_{max}$ $P_2 | reasgn(last(1)^*), Sum | C_{max}$ & ($1.41, 1.25$) Tight for respective setups . \\
\hline
Cao et al. 2012 [74] & $P_2 | Opt, Max | C_{max}$ $P_2 | B(k), Max | C_{max}$ $P_2 | B(1), Decr | C_{max}$ $P_2 | B(1), TGRP | C_{max}$ & ($1.2, 1.25, 1.16, 1.33$) Tight for respective setups. \\
\hline
Albers and Hellwig 2012 [75] & $P_m | Sum | C_{max}$ & $1.585$ LB,  $1.75$ UB\\
\hline 
Lan et al. 2012 [76] & $P_m | B(1.5m) | C_{max}$ & $1.5$ Tight \\
\hline
Cheng et al. 2012 [78] & $P_m | Decr | C_{max}$ & $1.18$ Tight for $m=3$, $1.25$ Tight for $m > 3$. \\
\hline
Lee and Lim 2013 [79] & $P_m |  Sum | C_{max}$ $P_m |  Max | C_{max}$ $P_m |  Sum, Max | C_{max}$ & ($1.4$, $1.4615$, $1.5$) UB for $m$=$3$,$4$,$5$  with \textit{Sum}, ($1.618$, $1.667$) Tight for $m$=$4$,$5$ with \textit{Max},  ($1.462, 1.5$) UB for $m$=$4$,$5$ with \textit{Sum} and \textit{Max}.\\
\hline
Kellerer and Kotov 2013 [104] & $P_m|Opt|C_{max}$ & $1.571$ UB.\\
\hline
Cao and Wan 2014 [84] & $P_2 | TGRP[1, b], Decr | C_{max}$ $P_2 | TGRP[ub], Decr | C_{max}$ & $1.16$ Tight for $TGRP[1, b]$ and $1 \leq b < 1.5$, $1.166$ LB for $TGRP[ub]$ and $ub \geq 1.5$. \\
\hline
Tang and Nie 2015 [87] & $P_m | Incr-r, Decr | C_{max}$ & ($\frac{3}{2} - \frac{1}{2m}$) UB. \\
\hline
Kellerer et al. 2015 [90] & $P_m | Sum | C_{max}$ & $1.585$ Tight for $m\rightarrow \infty$.\\
\hline
Gabay et al. 2015 [105] & $P_m|Opt|C_{max}$ & $1.5294$ UB.\\
\hline
Bohm et al. 2017 [106] & $P_m|Opt|C_{max}$  $P_3|Opt|C_{max}$ & $1.5$ UB for $P_m$, $1.375$ UB for $P_3$.\\
\hline
Gabay et al. 2017 [107] & $P_m|Opt|C_{max}$ & $1.357$ LB.\\
\hline
\end{tabular}
\label{tab: Important Results for Identical Machines: 2011-2017}
\end{table} 
\section{Emergence of Semi-online Scheduling Setups and Classification of the Related Works}\label{sec:Emergence of Semi-online Scheduling Setups and Classification of the Related Works}
\begin{figure}[h]
\centering
\includegraphics[scale=0.8]{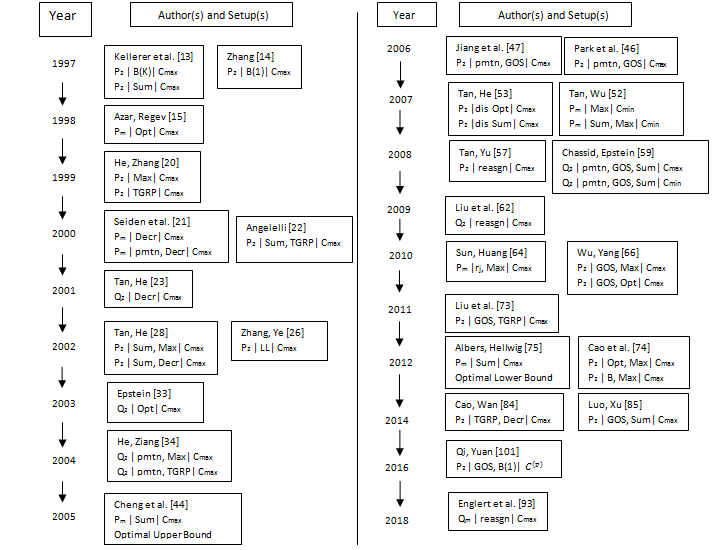}
\caption{Evolution Timeline of Semi-online Scheduling Setups}
\label{fig:researchtrends.png}  
\end{figure}

\begin{figure}[h]
\centering
\includegraphics[scale=0.86]{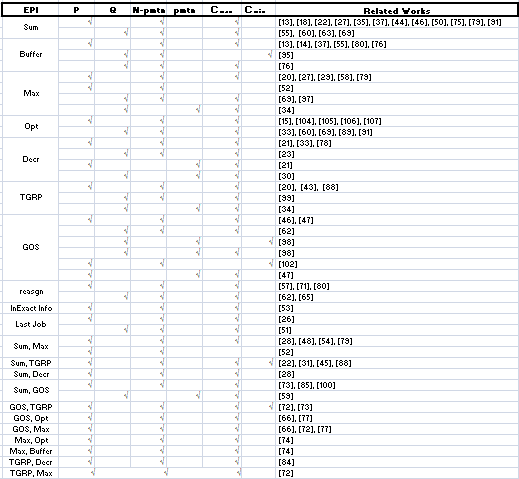}
\caption{Classification of Related Works based on EPI}
\label{fig:Relatedpapers.png}  
\end{figure}

\textbf{Evolution Time-line for Semi-online Scheduling Setups.} After making a comprehensive literature survey, we understand and explore various problem setups, research directions and research trends in semi-online scheduling. In this section, we sketch a time-line to represent the emergence of various semi-online scheduling setups as shown in figure \ref{fig:researchtrends.png}. \\
\textbf{Classification of Related Works based on EPI.}
Though many researchers have exhaustively studied semi-online scheduling based on either a single \textit{EPI} or more than one \textit{EPI}s, there is hardly any attempt to develop a taxonomy to classify the literature and related works based on \textit{EPI}. Here we attempt to classify the whole literature on semi-online scheduling based on \textit{EPI}s for identifying related works for various setups. We present our classification in figure \ref{fig:Relatedpapers.png}. We consider the setups for identical($P$) and uniform($Q$) machines in our classification. The other parameters for various problem setups are processing formats such as non-preemptive($N-pmtn$) and preemptive($pmtn$) and  optimality criteria such as makespan($C_{max}$) and \textit{Max-Min}($C_{min}$). We also provide links to references of related works for each problem setup.  For each \textit{EPI}, we have mentioned the various setups which are studied in the literature along with their references. Our classification may help the researchers to focus on related works and explore specific research directions for future work. The future research work can also be carried out based on specific \textit{EPI} by choosing a particular set up from our classification.

\section{Research Challenges and Open Problems}\label{sec:Research Challenges and Open Problems}
Semi-online scheduling has been extensively studied in various setups for the last two decades. Still, there are many research issues, which can lead to further investigations in this area. We conclude our survey on the important results and critical ideas for semi-online scheduling by exploring some of the non-trivial research challenges and open problems as follows.   
\subsection{Research Challenges}\label{subsec:Research Challenges}
\begin{itemize}  
\item Exploration of practically significant new \textit{EPI}s that can help in improving the \textit{CR} of the existing online scheduling algorithms. 
\item Generation, characterization and classification of the input job sequences that can resemble the real world inputs in various semi-online scheduling setups. 
\item Minimize or diminish the gap between \textit{LB} $1.585$ and \textit{UB} $1.6$ for the setup \textbf{$P_m|Sum|C_{max}$}. 
\item Reduce the \textit{CR} $1.366$ for preemptive \textbf{$P_m|C_{max}$} setting. 
\item Improvement of $1.5$-competitive strategy for \textbf{$P_2|C_{max}$} problem with inexact partial information. The solution for \textbf{$P_m|C_{max}$} problem is unknown in this case.
\item Close or remove the gap [$1.442, 1.5$] between lower and upper bound for semi-online scheduling on unbounded parallel batch machine. 
\item  Design of optimal semi-online scheduling algorithm with at most $1.5$-competitiveness for scheduling on $m$-identical machines ($m\geq 2$) under \textit{GOS} and known \textit{Opt}. 
\item Exploration of optimal semi-online algorithms for \textbf{$P_m|C_{max}$} and \textbf{$Q_m|C_{max}$} setups with reassignment of job policy.  \end{itemize} 
\subsection{Open Problems}\label{subsec:Open Problems} \begin{itemize} 
\item Can \textit{EPI} be used to develop a new complexity class for evaluating the performance of online algorithms?
\item Does there exist an optimal semi-online algorithm with \textit{CR} less than $1.33$ for the setup $P_m|B(k)|C_{max}$ with $k$=$1$?
\item Can the \textit{CR} $1.2$ be improved for the setup $P_2|Sum|C_{max}$? Can a matching bound be possible for $P_m|Sum|C_{max}$? How far preemption can help in improving the results in this setting?
\item How can we establish a relationship between semi-online scheduling and online scheduling with look ahead? Which one is practically significant? For an instance, is it possible to obtain a \textit{CR} less than or equal to $1.33$ for $P_m|Sum|C_{max}$, where $T$ is known for $k$ future jobs and $1\leq k < n$.
\item Can a tight bound be possible for $P_3|C_{min}$, which is independent of number of machines?
\item Does there exist an optimal semi-online algorithm for the setup $Q_m|C_{min}$? Already, a $1$-competitive algorithm is known for $Q_2|C_{min}$ due to [59].
\item What can be an optimal semi-online policy for $P_m|C_{max}$? Can a $1$-competitive semi-online algorithm be possible for this setting with known \textit{Decr} [108]?
\item What can be a tight bound for semi-online scheduling on uniform machines with overall speed ratio interval of $[1, \infty)$?
\item Does there exist an optimal semi-online strategy for multiple unbounded parallel batch processors?
\item Can \textit{EPI} be helpful in improving the best competitive bounds obtained for online scheduling in various setups of unrelated parallel machines? 
\end{itemize}  
\end{document}